\documentclass[12pt,aps,prd,preprint,tightenlines,superscriptaddress,showpacs,nofootinbib]{revtex4-1}
\pdfoutput=1
\usepackage{amsmath,amssymb,amsthm,amsfonts}
\allowdisplaybreaks
\usepackage{slashed} 
\usepackage{graphicx}
\usepackage{subfigure}
\usepackage{dcolumn}
\usepackage{hyperref}      
\usepackage{bm}
\usepackage{epsfig}
\usepackage{epstopdf}
\usepackage{setspace}
\usepackage{color}
\usepackage{slashed}
\newcommand{\Eqref}[1]{Eq.~(\ref{#1})}

\newcommand{\secref}[1]{Sec.~\ref{sec:#1}}

\newcommand{\appref}[1]{Appendix~\ref{sec:#1}}
\newcommand{\figref}[1]{Fig.~\ref{fig:#1}}

\newcommand{\tableref}[1]{Table~\ref{#1}}

\newcommand{\beq}{\begin{equation}}
\newcommand{\eeq}{\end{equation}}
\newcommand{\beqa}{\begin{eqnarray}}
\newcommand{\eeqa}{\end{eqnarray}}

\begin{document}

\title{Dark sectors and enhanced $h\to \tau \mu$ transitions }
\author{Iftah Galon}
\email{iftachg@uci.edu}
\affiliation{Department of Physics and Astronomy, University of
California, Irvine, CA 92697, USA}

\author{Jure Zupan}
\email{zupanje@ucmail.uc.edu}
\affiliation{Department of Physics, University of Cincinnati, Cincinnati, Ohio 45221,USA}
\affiliation{CERN, Theory Division, CH-1211 Geneva 23, Switzerland}

\preprint{UCI-HEP-TR-2016-17}
\preprint{MITP/16-114}
\preprint{CERN-TH-2016-260}

\begin{abstract}
LHC searches with $\tau$ leptons in the final state are always inclusive in missing-energy sources. 
A signal in the flavor-violating Higgs decay search, $h\to\tau\mu$, could therefore equally well be due to a flavor conserving decay, but with an extended decay topology with additional invisible particles.
We demonstrate this with the three-body decay $h\to\tau\mu\varphi$, where $\varphi$ is a flavorful mediator decaying to a dark-sector.
This scenario can give thermal relic dark matter that carries lepton flavor charges, a realistic structure of the charged lepton masses, and explain the anomalous magnetic moment of the muon, $(g-2)\mu$, while simultaneously obey all indirect constraints from flavor-changing neutral currents. 
Another potentially observable consequence is the broadening of the collinear mass distributions in the $h\to \tau\mu$ searches.
\end{abstract}

\vskip1cm
\maketitle

\section{Introduction}
The quark and lepton masses in the SM are highly hierarchical, with the electron roughly $\sim10^5$ times lighter than the top quark. It is possible that this hierarchy has a dynamical origin, and is due to a breaking of a horizontal flavor symmetry \cite{Froggatt:1978nt}. Rare Higgs decays are a natural place to search for a signal of such a possibility. First of all, the Higgs Yukawa couplings are directly tied to the generation of fermion masses. Secondly, the SM Higgs decay width is small, $\Gamma_{\rm SM}\simeq 4$~MeV, so that even feeble couplings of new states to the Higgs can have visible effects. In the SM all the Higgs decays are flavor diagonal, with $h\to b\bar b$ the dominant decay mode, followed by $h\to WW^*, gg, \tau\tau, \dots$ Nontrivial flavor dynamics, accompanied by new sources of electroweak symmetry breaking, can  lead to flavor violating decays such as, e.g., $h\to b\bar s$ or $h\to \tau \mu$ \cite{Altmannshofer:2015esa,Altmannshofer:2016zrn}. A discovery of such a decay would be a clear signal of New Physics (NP). 

In this paper we explore the possibility that the dark sector is charged under the same horizontal flavor symmetry as the SM fields. If the dark sector contains states lighter than the Higgs, this can have important consequences for the Higgs phenomenology. 
As a concrete example consider an extra light scalar from the dark sector, $\varphi$, with a horizontal charge such that the 
higher dimensional operators
\beq
\frac{1}{\Lambda} \frac{h}{\sqrt2}\bar \tau_L \mu_R \varphi,\quad {\rm or}\quad \frac{1}{\Lambda} \frac{h}{\sqrt2}\bar \mu_L \tau_R 
\varphi^*,
\eeq
carry no flavor suppression.
These operators lead to the exotic Higgs three-body decay
$h\to\tau\mu\varphi$. It is useful to compare its branching ratio with the one for the dominant Higgs decays to leptons,  $h\to \tau\tau$, 
\beq
\frac{Br(h\to\tau^\pm\mu^\mp\varphi~/\varphi^*)}
{Br(h\to \tau^+\tau^-)}
\simeq 
\frac{1}{6}\left(\frac{m_h}{4\pi \Lambda y_\tau} \right)^2
=0.66\times \Big(\frac{500 {\rm GeV}}{\Lambda}\Big)^2  \Big(\frac{0.01}{y_\tau}\Big)^2,
\eeq
where we follow the notation in \appref{analytic_expression}.
The $h\to\tau\mu\varphi$ branching ratio can thus be comparable to the one for
the two-body decay $h\to \tau\tau$, if NP resides at the TeV scale.
The relatively small tau Yukawa,
$y_\tau\simeq 1 \cdot 10^{-2}$, gives roughly the same suppression as the combination of phase-space and $\Lambda\sim {\mathcal O}(1~\rm{TeV)}$ suppression for the three-body decay.

The complex scalar $\varphi$ is assumed to primarily decay to a dark sector and
acts as an additional source of missing-energy in the event. The $h\to\tau\mu\varphi$ decay at the LHC would then be quite effectively captured by the present experimental $h\to\tau\mu$ analyses, depending on the details of the analysis and the decay kinematics. As we will show below, the hints in the
$h\to\mu\tau$  searches, 
\beq
\begin{split}
&{\rm CMS:}~~~~~Br(h\to \tau\mu)=(0.89\pm0.39)\% \text{~\cite{Khachatryan:2015kon}},
\\
&{\rm ATLAS}: 
~Br(h\to \tau \mu)=(0.53\pm 0.51)\%  \text{~\cite{Aad:2015gha,Aad:2016blu}},
\end{split}
\eeq
could in fact be entirely due to the $h\to\tau\mu\varphi$ decays (the 13 TeV CMS measurement~\cite{CMS:2016qvi} was not yet sensitive to the above branching ratios). An interesting question is then how one can distinguish between the two-body, flavor violating, $h\to\tau\mu$ decays and the three-body, flavor conserving, decays $h\to\tau\mu\varphi$. 

This paper is structured as follows. 
In \secref{model} we present the flavorful portal to DM model. We briefly review the use of U(1) horizontal symmetries, and apply them to generate the appropriate flavor structure of the lepton mass matrix and the mediator couplings. 
In Sec. \ref{sec:collider}, we perform a collider study for this model,
and analyze the parameter-space of couplings and masses that can
account for the observed $h\to \tau \mu$ CMS signal.
In \secref{bounds} we collect the constraints on the model from low energy precision measurements, 
while in \secref{darksec} we discuss the phenomenology of the flavorful dark sector.
Conclusions are given in \secref{conclusions}, while
\appref{analytic_expression} provides further details on our calculations of flavor changing transitions.

\section{Flavorful portal to dark matter}
\label{sec:model}
\subsection{Preliminaries}
We consider a model in which a dark sector interacts
with the SM leptons via a complex scalar field mediator, $\varphi$, 
a singlet under the SM gauge group.
The interactions of $\varphi$ with the visible sector are given by
dimension-five operators
\beq\label{eq:Lmed}
{\cal L}_{\rm vis-med.} \supset
  \frac{c_{ij}}{\Lambda}\bar L_i H E_j \varphi 
 + \frac{c'_{ij}}{\Lambda}\bar L_i H E_j \varphi^*
+ {\rm h.c.}.
\eeq
Here $H$ is the SM Higgs, $L_i, E_j$ are the SM lepton doublets and singlets, respectively, and $i,j=1,2,3,$ the generational indices. 
The suppression scale, $\Lambda$, arises from integrating vector-like fermions with masses ${\mathcal O}(1{\rm\,\,TeV})$. 
Additionally, $\varphi$ couples to the dark sector which contains two $Z_2$ odd fermions, $\chi_1$, and $\chi_2$, the lightest of which is a DM candidate. 
The interactions of $\varphi$ with the dark sector are given by the renormalizable operators
\beq
{\cal L}_{\rm dark} 
\supset
g^L_{ab}  \varphi \,\bar\chi_a P_L \chi_b  +g^R_{ab}\varphi\,\bar \chi_a P_R \chi_b + {\rm h.c.}, \qquad a,b=1,2.
 \label{eq:model}
\eeq

We pursue the idea  that 
an underlying theory of flavor governs all the flavor dependent couplings in the model.
This theory is responsible for generating the known hierarchy of lepton masses through the Yukawa matrix, $Y_{ij}^\ell$,
\beq
{\cal L}_{\rm vis.} \supset -Y^\ell_{ij} \bar L_i H E_j+  {\rm h.c.},
\label{eq:lepton_yukawa}
 \eeq
as well as the couplings of $\varphi$ to leptons, 
$c_{ij}$, $c_{ij}'$, and to the dark sector, $g_{ab}^L, g_{ab}^R$. We are interested in a flavorful dark sector \cite{
Kile:2011mn, 
Batell:2011tc, 
Kamenik:2011nb, 
Agrawal:2011ze, 
Lopez-Honorez:2013wla, 
Batell:2013zwa, 
Kile:2013ola, 
Kumar:2013hfa, 
Agrawal:2014una, 
Agrawal:2014aoa, 
D'Hondt:2015jbs, 
Agrawal:2015kje, 
Chen:2015jkt, 
Bhattacharya:2015xha, 
Bishara:2015mha, 
Agrawal:2015tfa, 
Haisch:2015ioa, 
Calibbi:2015sfa, 
Kilic:2015vka} 
where both the mediator, $\varphi$, as well as the DM fields, $\chi_{1,2}$, carry nonzero flavor charges. Phenomenologically very interesting is the situation where flavor dynamics generates a
single dominant off-diagonal coupling in the
$c_{ij},~c'_{ij}$ matrices, while all the others are suppressed.
We will be interested in the case where DM is lighter than the mediator, $m_\varphi> m_{\chi_1}$, so that the mediator can decay into the dark sector.
 We base our discussion on a concrete realization using the Froggatt-Nielsen construction \cite{Froggatt:1978nt}, though our main conclusions are more general. 

Recently, a
similar construction, but with $m_\chi > m_\varphi$,  was proposed in~\cite{Galon:2016bka}.
In this scenario dark matter
annihilates into on-shell mediators which subsequently decay
to opposite-charge different-flavor pairs of leptons.
Such annihilations can account for the excess observed by the FERMI-LAT collaboration in the spectrum of gamma-rays from the galactic center~\cite{TheFermi-LAT:2015kwa}. They
result in a softer $e^\pm$ energy spectrum then one obtains for flavor conserving interactions, and thus avoid the AMS-02 bounds~\cite{Aguilar:2014mma}.

A crucial assumption in these models is
that in the lepton mass basis 
\Eqref{eq:Lmed} contains only a {\it single} dominant mediator coupling, $c_{ij}$ or $c'_{ij}$, with all the other couplings suppressed.
In addition, the scalar potential for $\varphi$ is assumed to only
contain terms proportional to $|\varphi|^2$.
The single non-negligible coupling breaks the
lepton number symmetry $U(1)_i\otimes U(1)_j$ down to a
$U(1)_{i-j}$, under which $\varphi$ has a charge of $2$.
This residual symmetry is only
approximate. However, its breaking is small, 
so that the flavor structure is stable under the renormalization group.
As a result, the model does 
not lead to hazardously large Lepton Flavor Violating (LFV) transitions.

\subsection{The Froggatt-Nielsen Mechanism and Higher-Dimensional Operators}
In the Froggatt-Nielsen mechanism the fermion mass hierarchy and mixings are generated from broken $U(1)$ flavor symmetries, such
that the entries of the fermion mass matrix correspond to higher-dimension operators.
The smaller an entry is, the larger is the
dimension of its corresponding operator. 
 
In the phenomenologically realistic example we will use a flavor symmetry that is a product of two $U(1)$'s, $U(1)_A\times U(1)_B$. As a warm-up, however, we review the Froggatt-Nielsen mechanism for a single $U(1)$. In that case the  SM charged lepton Yukawa couplings $Y^\ell_{ij}$, Eq.~\eqref{eq:lepton_yukawa}, arise from 
\beq\label{eq:FN_effective_op}
{\cal L}_{\rm vis.} \supset - \alpha_{ij} \bar L_i H E_j \left(\frac{S~{\rm or}~S^*}{M}\right)^{|n^{Y}_{ij}|}.
\eeq
Here $\alpha_{ij}\sim {\cal O}(1)$ are flavor anarchic complex coefficients,
$S$ is a scalar field with flavor $U(1)$ charge $[S]_Q=-1$, while
$n^{Y}_{ij}=[\bar L_i]_Q + [E_j]_Q + [H]_Q$
is the sum of the flavor symmetry charges.
Whether $S$ or $S^*$ appear in \eqref{eq:FN_effective_op} depends on the sign of $n^{Y}_{ij}$. The mass scale $M$ is associated with heavy fermions that were integrated out and roughly coincide with the scale at which flavor is broken by the vev of $S$, 
\beq
\lambda = \frac{\langle S \rangle}{M} \simeq 0.2.
\eeq
The value of $\lambda$ is chosen to be close in size to the Cabibbo angle in order to reproduce the CKM matrix in the quark sector. The SM Yukawa for the charged leptons are thus given by 
\beq\label{eq:Yellij:FN}
Y^{\ell}_{ij} = \alpha_{ij} \lambda^{|n_{ij}^Y|}.
\eeq

The flavor structures of $c_{ij}$ and $c_{ij}'$ in \eqref{eq:Lmed} are generated in a similar way from   
\beq\label{eq:Lmed:beta}
{\cal L}_{\rm med.}\supset \beta_{ij} \bar L_i H E_j \left(\frac{S~{\rm or}~S^*}{M}\right)^{|n^{c}_{ij}| }
\frac{\varphi}{\Lambda}+\beta_{ij}' \bar L_i H E_j \left(\frac{S~{\rm or}~S^*}{M}\right)^{|n^{c'}_{ij}| }
\frac{\varphi^*}{\Lambda},
\eeq
where $\beta_{ij}\sim\beta_{ij}'\sim{\cal O}(1)$ are unknown coefficients, and $n^{c(c')}_{ij} = [\bar L_i]_Q + [E_j]_Q + [H]_Q \pm [\varphi]_Q$. For simplicity we take $\Lambda\simeq M$,  but in general $\Lambda$ and $M$ are unrelated. After $S$ obtains a vev and breaks the flavor symmetry Eq.  \eqref{eq:Lmed:beta} gives
\beq\label{eq:cij:FN}
c_{ij} \sim \lambda^{|n^{c}_{ij}|}, \qquad c_{ij}' \sim \lambda^{|n^{c'}_{ij}|}.
\eeq
Here the similarity sign denotes equality up to ${\mathcal O}(1)$ coefficients.
 The couplings of $\varphi$ to dark sector fields, $g_{ab}^{L,R}$, are generated in an analogous way,
 \beq\label{eq:gab:FN}
 g_{ab}^{L}\sim g_{ab}^{R}\sim \lambda^{|n_{ab}|},
 \eeq
 where $n_{ab}=- [\chi_a]_Q+[ \chi_b]_Q+[ \varphi]_Q$.

The above results are easily generalized to the case of more than one $U(1)$ flavor symmetry. We find that a phenomenologically viable description is obtained for a product  of two $U(1)$ factors,
$U(1)_A\times U(1)_B$. 
Each is broken by a corresponding complex scalar field $S_{A,B}$ with the flavor charges 
\footnote{
The charge assignment of any field $\psi$ 
under $U(1)_A\times U(1)_B$ is 
denoted by $[\psi]_Q=\big([\psi]_{Q_A}, [\psi]_{Q_B}\big)$.
}
$[S_A]_Q= (-1,0)$ and $[S_B]_Q= (0,-1)$. For simplicity we take the vevs of $S_A$ and $S_B$ to be equal, as we do for the related mass scales $M_{A,B}$, so that
\beq
\frac{\langle S_A \rangle }{M_{A}} = \frac{\langle S_B \rangle }{M_{B}} =\lambda \simeq 0.2.
\eeq
The results for the $Y_{ij}^\ell$, $c_{ij}^{(')}$, and $g_{ab}^{L,R}$ flavor structures are obtained from \eqref{eq:Yellij:FN}, \eqref{eq:cij:FN}, \eqref{eq:gab:FN} by simply exchanging
\beq
\lambda^{|n_{ij}|}\to \lambda^{|n^{A}_{ij}| + |n^{B}_{ij}| },
\eeq
where $n^{A}_{ij}$ and $n^{B}_{ij}$ are the corresponding sums of
charges for $U(1)_A$ and $U(1)_B$, respectively.

\subsection{A Concrete Realization}
\label{sec:concrete}
As a concrete realization of a $U(1)_A\times U(1)_B$ flavor model we choose the following charge assignments for the SM leptons and the scalar field $\varphi$,  
\beqa
\begin{matrix}\label{eq:charge:assignments}
[\bar L_1]_Q&=& (7,1), & ~\qquad~ & [E_1]_Q&=& (-7,7), \\
[\bar L_2 ]_Q&=&(-6,-2), &~\qquad~ & [E_2]_Q&=& (6,-3), \\
[\bar L_3]_Q&=& (-2,-4), &~\qquad~ & [E_3]_Q&=& (1,6), \\
[H]_Q&=& (0,0), &~\qquad~ &[\varphi]_Q&=& (5,-4).
\end{matrix}
\label{eq:charge_assignment}
\eeqa
The flavor dependent couplings then have the following patterns,
\beq
\label{eq:coupling_pattern}
Y^{\ell} \sim
\begin{pmatrix}
\lambda^8 & \lambda^{15} & \lambda^{15} \\
\lambda^{18} & \lambda^{5} & \lambda^{9} \\
\lambda^{12} & \lambda^{11} & \lambda^{3}
\end{pmatrix}
,\quad
c \sim
\begin{pmatrix}
\lambda^{9} & \lambda^{24} & \lambda^{16} \\
\lambda^{9} & \lambda^{14} & 1 \\
\lambda^{5} & \lambda^{20} & \lambda^{6}
\end{pmatrix}
,\quad
c' \sim
\begin{pmatrix}
\lambda^{17} & \lambda^{10} & \lambda^{14} \\
\lambda^{27} & \lambda^{6} & \lambda^{18} \\
\lambda^{21} & \lambda^{4} & \lambda^{12}
\end{pmatrix}.
\eeq
These are consistent with the lepton mass eigenvalues
\beq
\{m_e,~m_\mu,~m_\tau\} \sim \frac{v}{\sqrt2}\{\lambda^8, \lambda^5, \lambda^3\},
\eeq
obtained by diagonalizing the charged lepton mass matrix with 
the left- and right- rotation matrices that scale as
\beq
V_{L_L} \sim
\begin{pmatrix}
1 & \lambda^{10} & \lambda^{12} \\
\lambda^{10} & 1 & \lambda^{6} \\
\lambda^{12} & \lambda^{6} & 1
\end{pmatrix}
,\qquad
V_{E_R} \sim
\begin{pmatrix}
1 & \lambda^{13} & \lambda^{9} \\
\lambda^{13} & 1 & \lambda^{8} \\
\lambda^{9} & \lambda^{8} & 1
\end{pmatrix}
\label{eq:mixing_pattern}.
\eeq
Note that the charge assignments in \Eqref{eq:charge:assignments} are large in order to get the $\sim \lambda^8$ suppression of the electron mass. Smaller flavor charges are possible in a two-Higgs doublet model, if $m_e,m_\mu$ masses do not come predominantly from the SM Higgs vev, but rather from a small vev of the heavier Higgs \cite{Altmannshofer:2015esa}. We do not pursue this possibility further. 

Generating the flavor structure using two $U(1)$s is advantageous
since $c'$ can be chosen to be aligned with the lepton mass basis. 
Indeed, \Eqref{eq:mixing_pattern} shows that the rotations of $Y^\ell$ to the lepton mass basis are highly suppressed,
and as a result, do not induce large couplings in $c,~c'$, 
other than the single dominant one.  Rotating to the mass, basis, the couplings read
\beq
\label{eq:coupling_pattern_mass_basis}
c \sim
\begin{pmatrix}
\lambda^{9} & \lambda^{18} & \lambda^{10} \\
\lambda^{9} & \lambda^{8} & 1 \\
\lambda^{5} & \lambda^{14} & \lambda^{6}
\end{pmatrix}
,\quad
c' \sim
\begin{pmatrix}
\lambda^{17} & \lambda^{10} & \lambda^{14	} \\
\lambda^{19} & \lambda^{6} & \lambda^{14} \\
\lambda^{17} & \lambda^{4} & \lambda^{12}
\end{pmatrix}.
\eeq
Below, we will discuss the viability of \Eqref{eq:coupling_pattern}
with respect to flavor observables, showing that one avoids all present constraints.

The phenomenology also crucially depends on the flavor structure of the dark sector. Taking $[\chi_1]_Q-[\chi_2]_Q = [\varphi]_Q$ one has 
\beq
g_{ab}^L\sim g_{ab}^R \sim 
\begin{pmatrix}
\lambda^9 & 1\\
\lambda^{18} & \lambda^9
\end{pmatrix},
\label{eq:chi_varphi_cpl}
\eeq
for the charge assignment of $\varphi$ in \eqref{eq:charge:assignments}.  We see that the $\varphi$ coupling to the dark sector is maximally flavor violating with $\varphi$ coupling to $\bar \chi_1 \chi_2$  much larger than the remaining couplings, while the DM mass matrix is already almost completely diagonal in the basis of \eqref{eq:chi_varphi_cpl}.
The two dominant $\varphi$ decay modes are thus $\varphi\to \bar \chi_1 \chi_2$ and $\varphi\to \bar \tau_R \mu_L$ with the corresponding branching ratios given by
\beq
\frac
{
Br(\varphi\to \bar \tau \mu)
}
{
Br(\varphi\to \bar\chi_2 \chi_1)
}
\simeq
\frac12 \Big(\frac{ v}{\Lambda}\Big)^2 \frac{|c_{23}|^2}
{|g_{12}^L|^2+|g_{12}^R|^2},
\eeq
neglecting the masses of final state particles.
For $c_{23} \sim g_{12}^L\sim g_{12}^R$, and $\Lambda\sim{\mathcal O}(\rm{TeV})$, the mediator $\varphi$ primarily decays to the dark sector. The production of $\varphi$ then leads primarily to missing-energy signatures. 
We explore this scenario in the next section and show
its consequences for the LHC searches.
In the opposite limit, $c_{23} \gg g_{12}^L,g_{12}^R$, the $\varphi \to \bar \tau \mu$ decay mode can be the dominant decay mode. More precisely, this occurs if $[\varphi]_Q$ is significantly different from $[\chi_1]_a-[\chi_2]_b$ for all $a,b$, so that $g_{ab}^L,g_{ab}^R \ll 1$. 
In that case, the $\varphi\to\tau\bar\mu$ decay mode dominates,
resulting in a new exotic Higgs decay signature 
$h\to \tau^+\tau^-\mu^+\mu^-$ 
where one opposite charge $\tau\mu$ pair comes 
from the decay of an intermediate $\varphi$ particle.

\begin{figure}[t]
\centering
\includegraphics[width=0.22\textwidth]{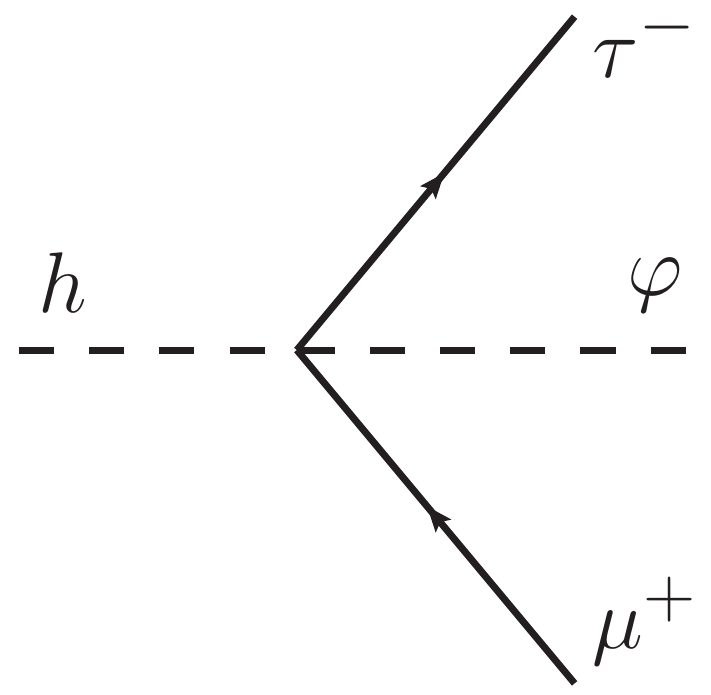}
\qquad\qquad\qquad
\includegraphics[width=0.22\textwidth]{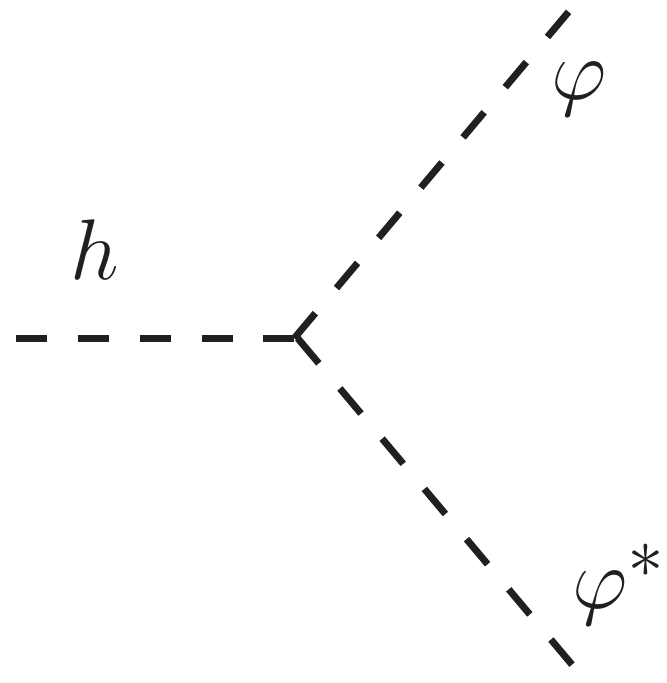}
\caption{The two new types of Higgs decays induced by the presence of the mediator $\varphi$. The $\varphi$ can decay either through $\varphi \to \tau \mu$ or $\varphi \to {\it invisible}$, depending on the details of the theory parameter space, see main text.}
\label{fig:modelI_diags}
\end{figure}

The  large flavor charge of $\varphi$ also implies that  the terms in the scalar potential $V(H,\varphi, \varphi^*)$ that have an odd number of $\varphi, \varphi^*$ fields are suppressed by the flavor symmetry. In this work we assume that $\varphi$ does not obtain a vev. The terms leading to $h-\varphi$ or $h-\varphi^*$ mixing are then suppressed by flavor symmetry and can be neglected in 
our analysis. 
The terms with no net flavor charges, such as the quartic terms
 $|\varphi|^4$ and $|H|^2|\varphi|^2$, are  expected to have 
${\cal O}(1)$ couplings.  The flavor violating couplings such as $\varphi\varphi$ are suppressed, in our case by $\lambda^{18}$. This means that the $\varphi$ field leads to two, almost degenerate mass eigenstates, $\varphi_{1,2}$, with relative mass splitting of ${\mathcal O}(\lambda^{18})$. The $|H|^2|\varphi|^2$ term leads to the
$h\to\varphi\varphi^*$ decay after electroweak symmetry breaking, see Fig.~\ref{fig:modelI_diags}. If $\varphi$ decays predominantly to the dark sector, the $h\to \varphi\varphi^*$ decays 
are constrained by the bound on the $h\to {\it invisible}$ branching ratio~\cite{Aad:2015pla,Aad:2015txa}. If the dominant decay of $\varphi$ is the $\varphi\to \bar \tau \mu$ channel, then the $h\to\varphi\varphi^*$ decay leads to the  $h\to (\tau^+ \mu^-)(\tau^- \mu^+)$  signature, where each of the $\tau\mu$ pairs originates from the $\varphi$ resonance.  In this scenario the $h\to 2\tau2\mu$ decay would thus have both the di-resonance and the three-body (single $\varphi$ resonance) contributions. Such exotic decays can be searched for by a modification of the flavor conserving di-resonance searches~\cite{Aad:2015oqa}.

One could relax our assumption that $\varphi$ has a vanishing vev. In that case \eqref{eq:Lmed} would contribute to the lepton mass matrix. In the limit when this is the dominant contribution to the lepton masses, both the Higgs Yukawa couplings as well as the Higgs couplings to $\varphi$ and leptons are governed by the same matrix. They are both diagonal in the charged lepton mass basis, while the Yukawa couplings are proportional to the charge lepton masses, as in the SM. More interesting is the case where \eqref{eq:Lmed} and \eqref{eq:lepton_yukawa} both give comparable contributions to the charged lepton mass matrix. In this case one would also need to include $h-\varphi$ mixing. To simplify our analysis we do not pursue this possibility further.

%

In the remainder of the paper we assume that $g_{ab}^{L,R}$ are given by Eq.~\eqref{eq:chi_varphi_cpl}, and that $\varphi\to{\it invisible}$ is the dominant decay mode. The $h\to \tau \mu\varphi$ thus appears in the detector as the $h\to \tau\mu$ decays with an additional missing-energy source. In the next section we explore whether or not such a decay could mimic in the experimental analysis the two-body $h\to \tau\mu$ decay, i.e., the decay with the same visible final state particles but without the additional missing-energy source.

\section{Collider Study}
\label{sec:collider}

We study the acceptance of the $h\to \tau \mu$ CMS search \cite{Khachatryan:2015kon} to the $h\to \tau \mu(\varphi\to {\it invisible})$ signal predicted in the model of Section \ref{sec:concrete}.
We implement the model in {\tt FeynRules}~\cite{Alloul:2013bka} and export it to {\tt MadGraph5\_aMC@NLO}~\cite{MG5}
in the UFO  format~\cite{Degrande:2011ua}. 
Using {\tt MadGraph} we generate the Higgs production through Gluon Gluon Fusion (GGF)~\cite{Alloul:2013naa} and through the Vector Boson Fusion (VBF)\footnote{
While the GGF contributions to the VBF optimized 2-jet categories are know to be small we still generate in {\tt MadGraph} 0,1, and 2 jet events, and apply the MLM matching scheme~\cite{Mangano:2006rw}.
},
as well as the subsequent $h\to \tau\mu \varphi$ decay.
Tau decays are simulated using {\tt TAUOLA}~\cite{Jadach:1990mz,Jezabek:1991qp,Jadach:1993hs}, while for parton-showering and hadronization we used 
{\tt Pythia 6}~\cite{PYTHIA6}. Detector simulation is performed using 
{\tt Delphes 3}~\cite{Delphes} with an internally implemented anti-$k_T$ jet algorithm~\cite{AntiKt} applying the {\tt FastJet}~\cite{Cacciari:2011ma} package.
We analyze the generated events in {\tt ROOT}~\cite{Brun:1997pa}, implementing the cut-flow of the CMS analysis  \cite{Khachatryan:2015kon}. 

A similar recast
of the ATLAS analyses~\cite{Aad:2015gha,Aad:2016blu} would be more involved because of the 
use of ``Missing Mass Calculator'' (MMC), a log-likelihood based
method for the reconstruction of taus in a hadron collider~\cite{Elagin:2010aw}. 
In the future it would be interesting to explore to what extent the MMC reduces the search acceptance to the invisible $\varphi$ in the decay, and whether the $h\to\tau\mu\varphi$ topology could explain the lack of signal in ATLAS.
In the leptonic channels ATLAS relies on the muon-electron momentum asymmetry of the $h\to \mu\tau_e$ decay.
While the muon and the tau share the Higgs momentum quite evenly, the electron shares the tau energy with neutrinos, and is therefore softer than the muon. In contrast, the SM background sources are highly symmetric under electron-muon exchange. The search, therefore applies a cut-flow which targets the two-body decay characteristics, and requires a muon of higher $p_T$ than the electron.
Such a strategy, nonetheless, reduces the sensitivity to the $h\to \tau\mu\varphi$ signal in which the $\tau$-$\mu$ pair is apriori not symmetric, because energy is also carried by $\varphi$, and the decay products are angularly denser than in a two-body decay.

\subsection{Analysis Results}
The CMS analysis \cite{Khachatryan:2015kon} 
divides signal into six categories;
the events with hadronic taus,  $\tau_h$, and the events with 
 taus decaying to electrons, $\tau_e$,  each of which are further split according to
the number of hard jets in the event,
$N_{Jets} = 0,1,2$.
After the cut flow in each signal category
the Signal Region (SR) is defined by
 \beq
100~\rm{GeV} \le m_{\tau\mu}^{(Coll)} \le 150~\rm{GeV},
\label{eq:SR_def}
\eeq
where $m_{\tau\mu}^{\rm (Coll)}$ is the invariant mass of the 
$\tau\mu$ pair in the ``collinear approximation''~\cite{Ellis:1987xu}.
In this approximation the net three momenta of invisible and visible final state particles in the $\tau$ decay
are assumed to be aligned. The collinear approximation works well
when the $\tau$ is highly boosted,
see~\cite{Elagin:2010aw, Bianchini:2014vza} for a more detailed discussion.
The collinear invariant mass is given
by
\beq\label{eq:mcoll}
m_{\tau\mu}^{\rm (Coll)} = 
\sqrt{
\left(
p_{\tau}^{\rm (vis)} + p_{\mu} + p_{\nu's}
\right)^2,
}
\eeq
where $p_\mu$ is the muon four momentum, and $p_{\tau}^{\rm (vis)}$
the four momentum of visible decay products, i.e., the electron momentum for $\tau_e$ and the momentum of $\tau$-tagged jet for $\tau_h$.
The proxy for the total neutrino four momentum, $p_{\nu's}$, is constructed by promoting
 the missing-energy transverse 2-vector, $\vec {\slashed E}_T$, 
to a massless 4-vector using
\beq
\vec p_{\nu's} =
\left(
 \vec {\slashed E}_T \cdot \hat p_{\tau^{(vis)}}.
 \right)
 \hat p_{\tau^{(vis)}}
\eeq
Here $\hat p_{\tau^{(vis)}}$ is the unit 3-vector in the direction of visible $\tau$ decay products.
Notably, this construction suggests that the search is inclusive in all missing-energy sources, including, but not restricting to, the neutrino decay products of the $\tau$.

For $h\to \tau \mu \varphi$ decays  we examine four benchmark $\varphi$ masses, 
$m_\varphi = 5, 10, 15, 20~\rm{GeV}$,
and compare the event yields to the CMS results.
We normalize our results in the GGF and the VBF channels to the LHC Higgs working group production cross-section at
$8~\rm{TeV}$, i.e., to $\sigma(gg\to h)_{\rm GGF}|_{j=0,1,2} = 19.27~\rm{pb}$, and to the NNLO QCD+NLO EW prediction $\sigma(pp\to h j j)_{\rm VBF} = 1.6~\rm{pb}$ ~\cite{Heinemeyer:2013tqa}, respectively.
We perform a simultaneous fit to the signal event yields in all six signal regions, 
using the reported
CMS yields and errors.
As a consistency check we also include 
the two-body $h\to\mu\tau$ decay topology.
Assuming the $h\to \tau\mu$ decay, the CMS collaboration obtained the best fit value for the corresponding Yukawa coupling 
$Y_{\tau\mu}=(3.7\pm0.8)\cdot 10^{-3}$ (setting the Yukawa for the other chirality structure to zero). 
This is in reasonable agreement with the best fit value from our recast, $Y_{\tau\mu}=2.4 \cdot 10^{-3}$, giving credence to our analysis.

%
\begin{table}[t]
\begin{tabular}{ccccc}
\hline\hline
Decay & $m_\varphi$ [GeV]& Br & Coupling  &
\\ \hline
$h\to\tau\mu$  & $-$  &~~ $3.6\times 10^{-3}$  ~~     & ~~ $Y_{23} = 2.4\times 10^{-3}$ ~~          &            
\\ 
$h\to\tau\mu\varphi$& $5$ &  $1.9\times 10^{-2}$     & $ c_{23} = 1.4$           &            
\\ 
$h\to\tau\mu\varphi$& $10$ & $2.6\times 10^{-2}$       &  $c_{23} = 1.7$           &            
\\
$h\to\tau\mu\varphi$& $15$ & $3.4\times 10^{-2}$      &  $c_{23} = 2.1$          &            
\\
$h\to\tau\mu\varphi$& $20$ & $4.8\times 10^{-2}$      &    $c_{23} =2.7$         &            
\\ \hline\hline
\end{tabular}
\caption{ Last column shows the best fit values for the couplings  $c_{23}$  for $h\to \tau\mu\varphi/\varphi^*$, setting $\Lambda=1$ TeV, and $Y_{23}$ for $h\to \tau\mu$, obtained from comparing our Monte Carlo study with the CMS excess. The third column gives the corresponding exotic Higgs decay branching ratio which is calculated using \Eqref{eq:higgs_2body}, and \Eqref{eq:higgs_3body}.
}
\label{table:results}
\end{table}

\begin{figure}[t]
\centering
\includegraphics[width=0.6\textwidth]{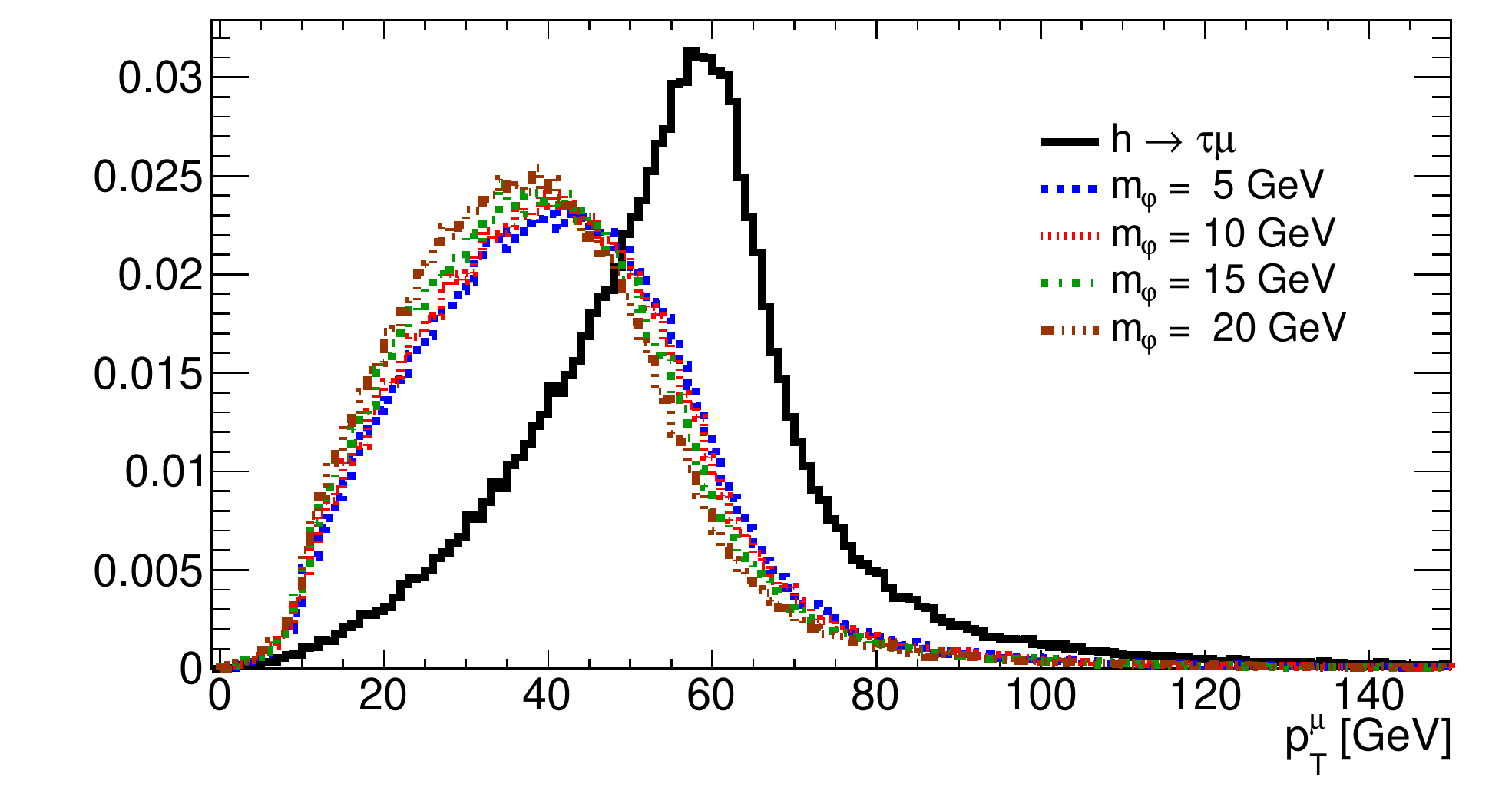}
\caption{
The normalized  distributions of the $h\to \tau\mu\varphi$ and $h\to \tau\mu$ signals as functions of the muon $p_T$. The color coding is
blue -- $m_\varphi = 5~\rm{GeV}$,
red -- $m_\varphi = 10~\rm{GeV}$,
magenta -- $m_\varphi = 15~\rm{GeV}$,
green -- $m_\varphi = 20~\rm{GeV}$.
}
\label{fig:ptmuon}
\end{figure}
%

In \tableref{table:results} we report
the best fit values for the Higgs decay branching fractions, 
and the best fit values for the corresponding couplings, $c_{23}$ for $h\to \tau \mu\varphi$ (setting $\Lambda=1$ TeV) and $Y_{23}$ for $h\to \tau\mu$,  that are required in order to explain
the CMS results. The comparison of the best fit value for $Y_{\tau\mu}$ in the recast to the CMS analysis indicates that the absolute values of the extracted couplings $c_{23}$ carry an ${\mathcal O}(1)$ uncertainty. A much smaller  uncertainty is expected, though, in the relative values of $c_{23}$ for different benchmarks, or in the ratios such as $c_{23}/Y_{\tau\mu}$.
 
From \tableref{table:results} we see that a three-body decay $h\to \tau\mu\varphi$ requires a factor of a few larger branching ratios to describe the data well than does the two-body $h\to \tau\mu$ decay. The reason is that the CMS search was optimized for a two-body decay, and thus only a subset of $h\to \tau\mu\varphi$ decays pass the cut-flow, resulting in a decreased signal acceptance. 
In particular, a three-body decay is much denser, and the particles are softer than in the two-body case. 
We show this in \figref{ptmuon} where we plot the muon $p_T$ distributions of the simulated models, and compare the spectrum 
of the $h\to \tau\mu$ decay to the the $h\to \tau\mu\varphi$ ones.
Indeed, muons are harder in the former case, and soften with increasing $\varphi$ mass. More importantly, while the $\tau$ has a roughly identical $p_T$-spectrum, for a given muon $p_T$, the $\tau$ $p_T$ is softer in the three-body case, and decreases as $m_\varphi$ grows. The acceptance thus decreases with $m_\varphi$, requiring increasing branching fractions to account for the observed result.

To explain the CMS excess one requires $c_{23}\sim {\mathcal O}(1)$ for electroweak scale $\Lambda\sim {\mathcal O}({\rm TeV})$, in agreement with the expectations from our flavor model in Section \ref{sec:concrete}.
Note that the inclusion of non-zero $c_{32}$, $c'_{23}$, or $c'_{32}$ would give identical collider phenomenology and only result in a trivial rescaling of the coupling constants.

\begin{figure}[t]
\centering
\includegraphics[width=0.48\textwidth]{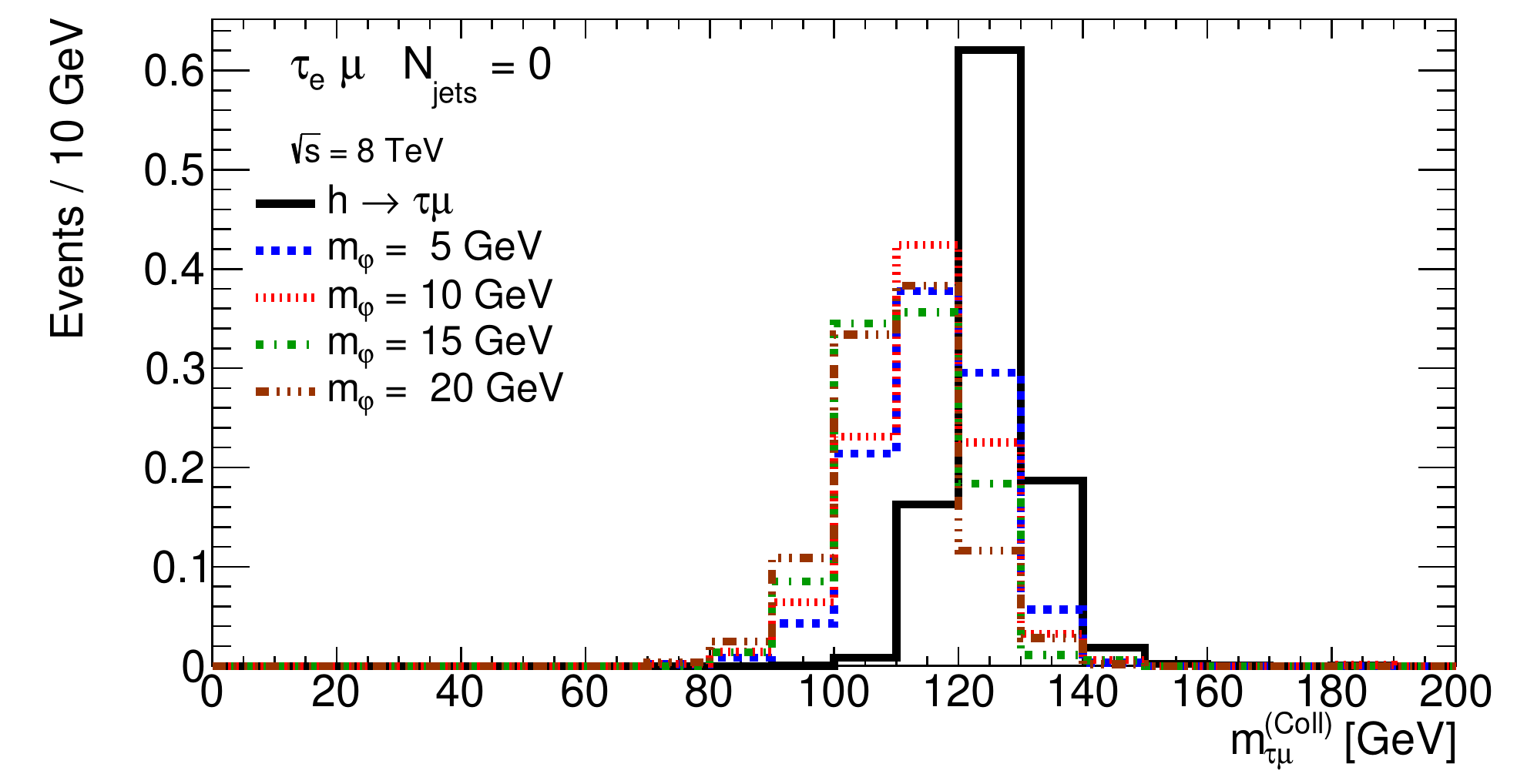}
~~
\includegraphics[width=0.48\textwidth]{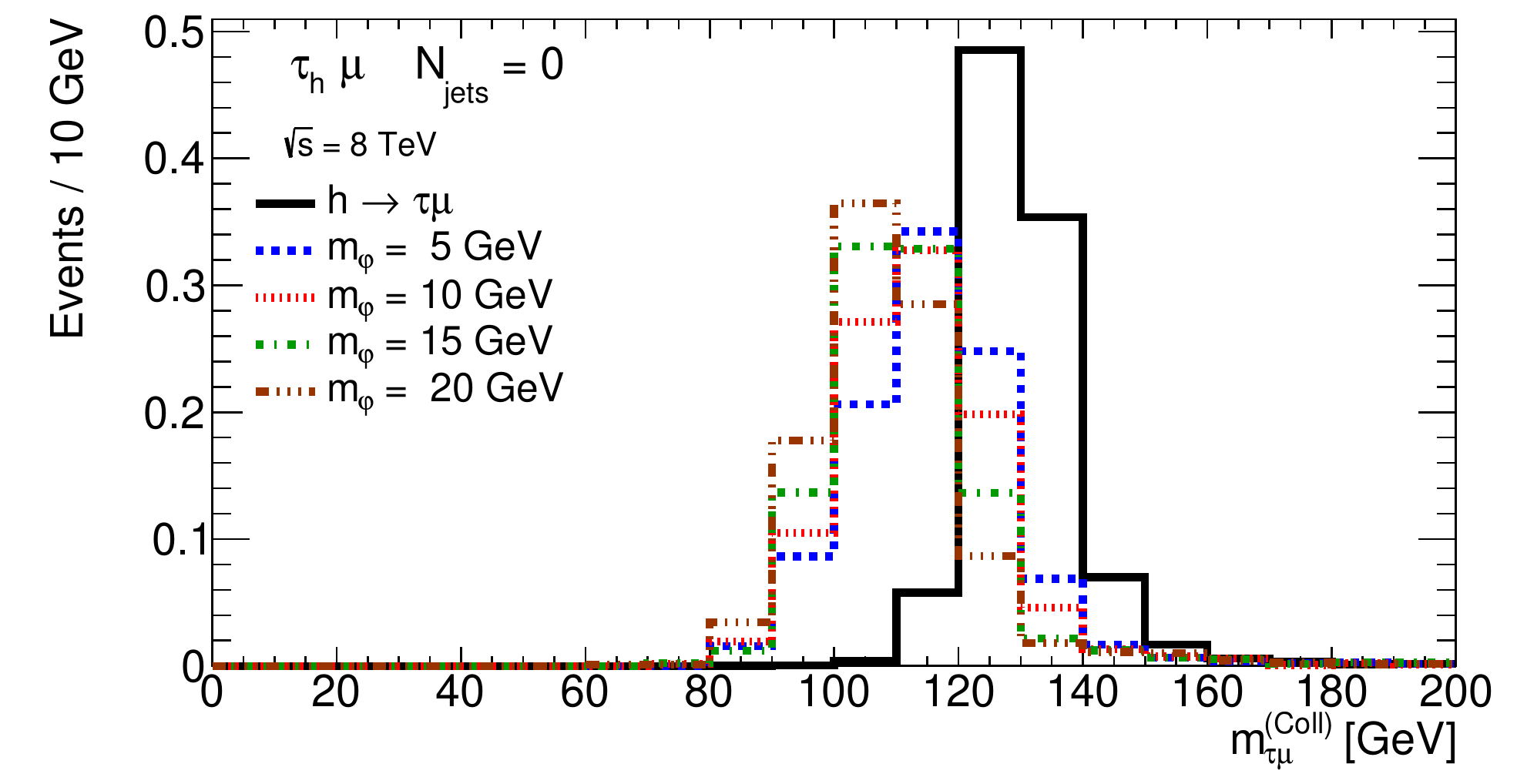}
\\
\includegraphics[width=0.48\textwidth]{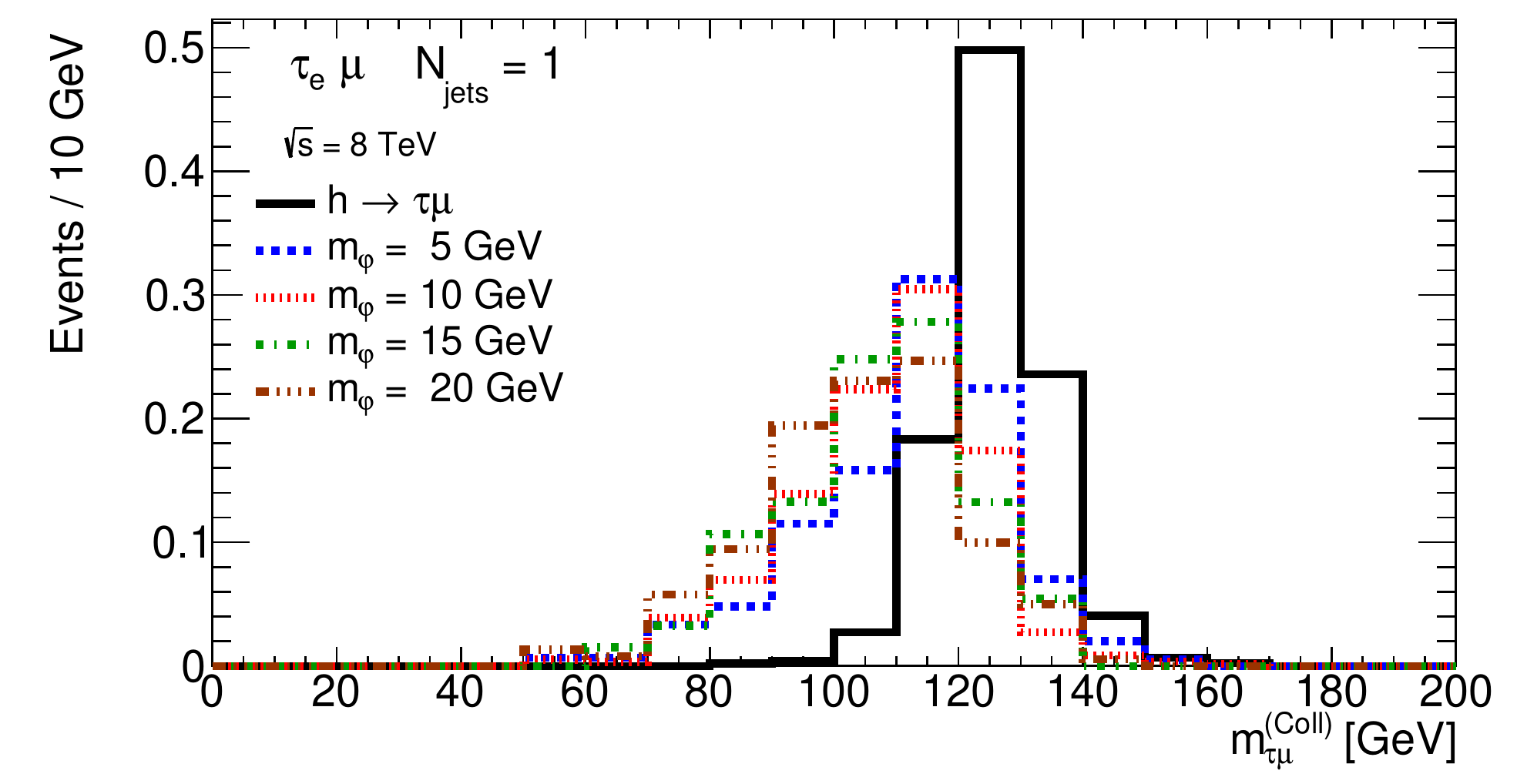}
~~
\includegraphics[width=0.48\textwidth]{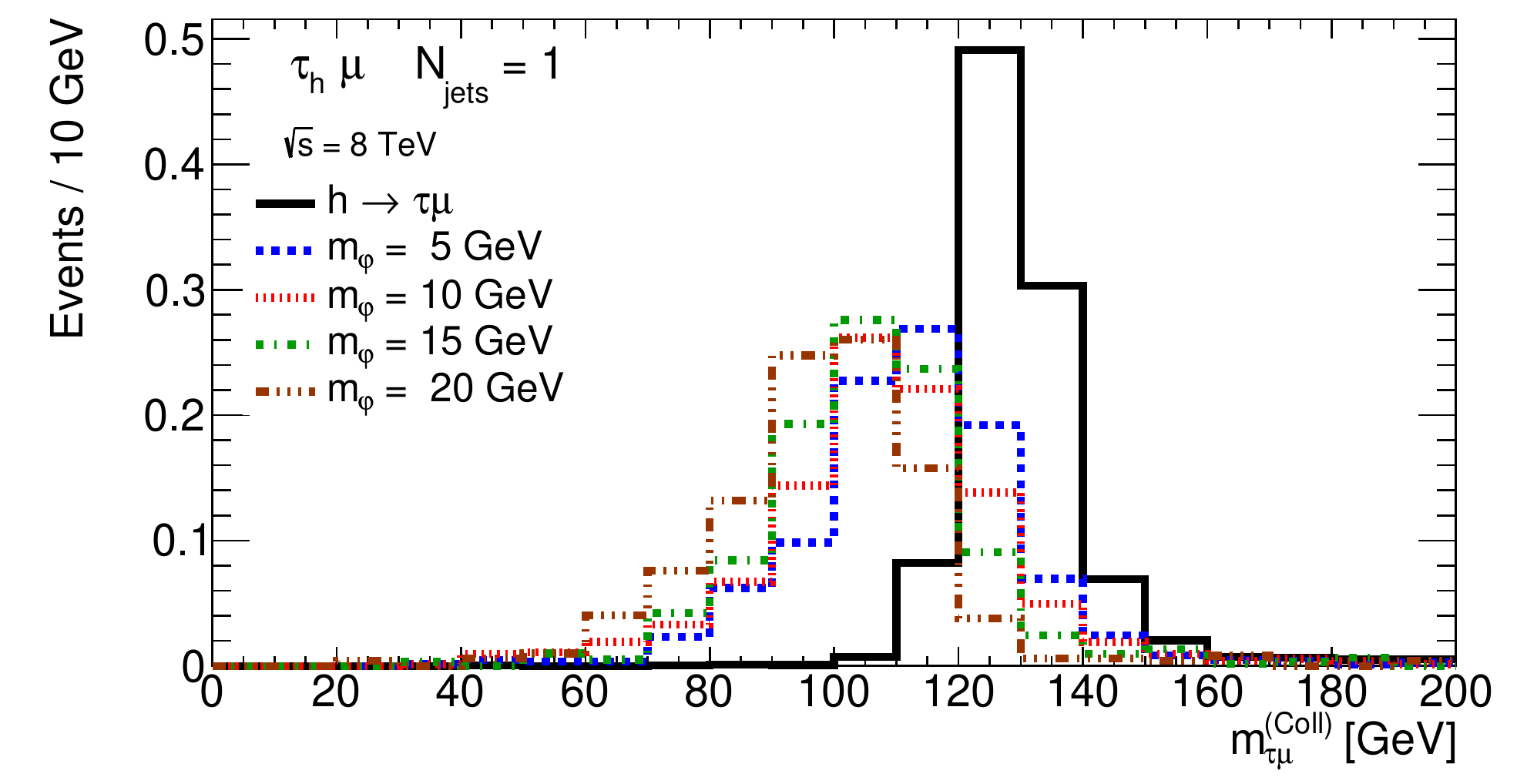}
\\
\includegraphics[width=0.48\textwidth]{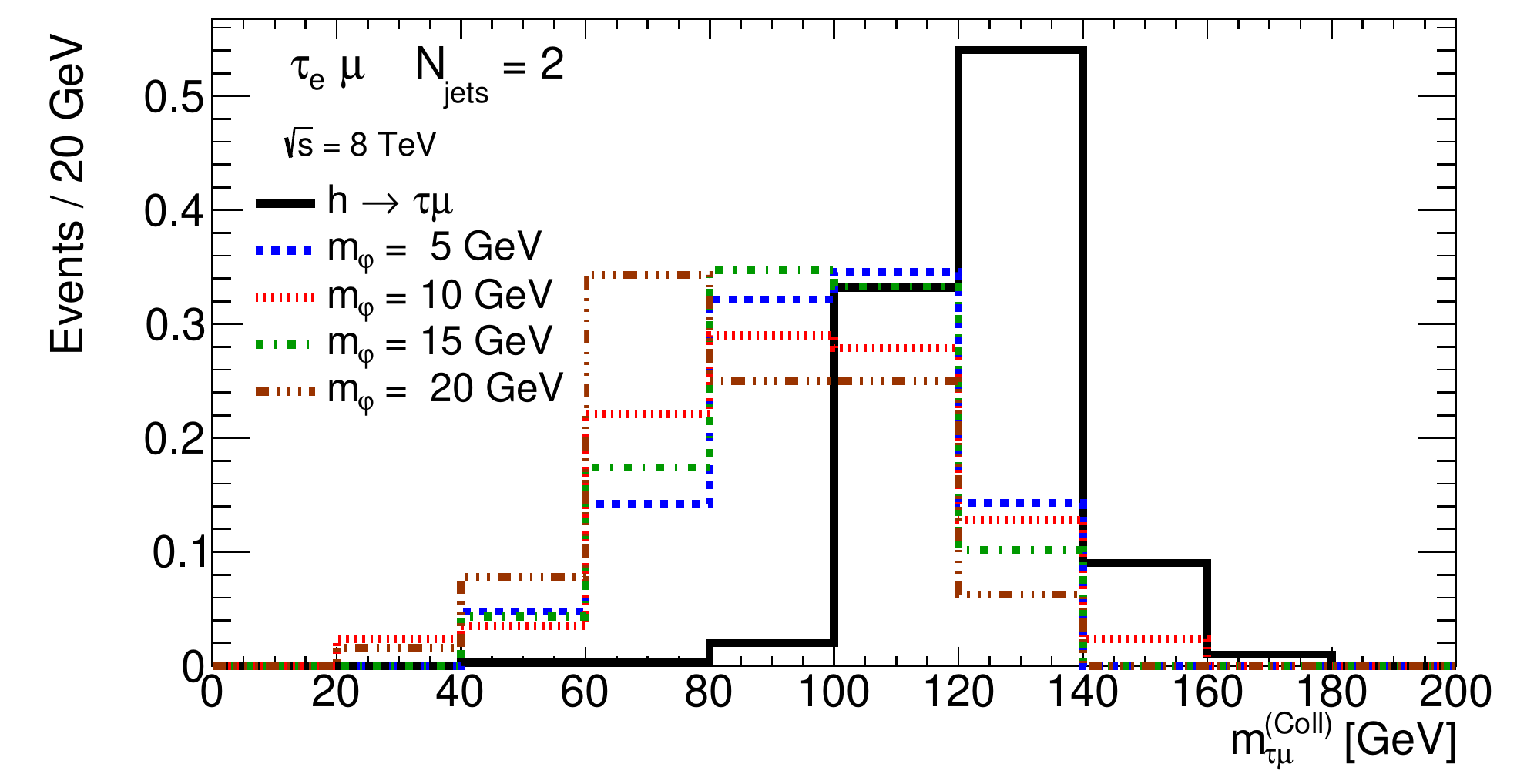}
~~
\includegraphics[width=0.48\textwidth]{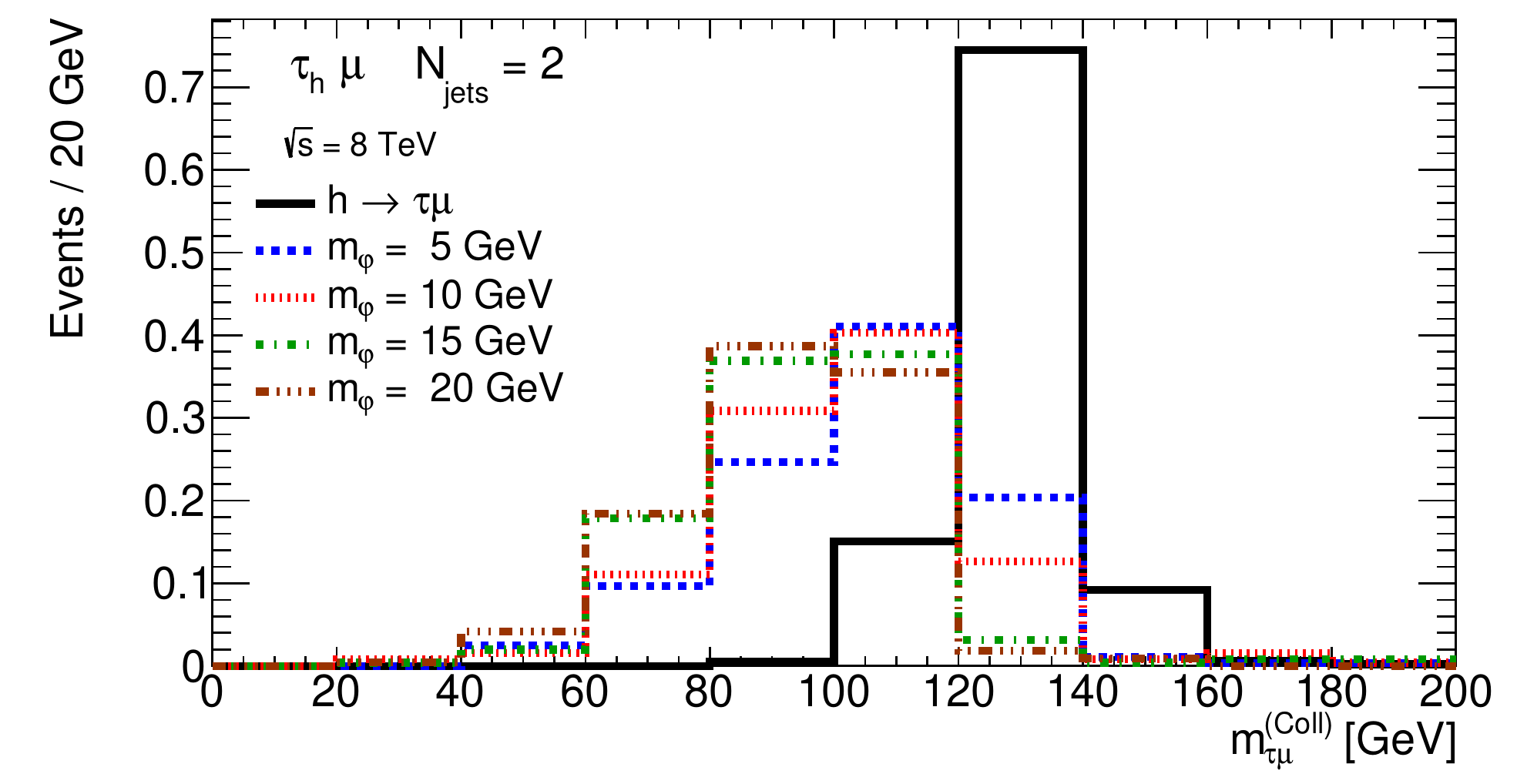}
\caption{
The normalized  distributions for 
the invariant mass of the $\tau\mu$ pair calculated
in the collinear approximation, $m_{\tau\mu}^{\rm (Coll)}$.
Left: Signal categories for the $\tau_e$ case, top to bottom: $0,1,2$-jets.
Right: Signal categories for the $\tau_h$ case, top to bottom: $0,1,2$-jets.
The black line denotes the normalized distribution for the $h\to\tau\mu$ decay,
while the other curves denote the $h\to\tau\mu\varphi$ decay benchmarks.
The color coding is as in Fig. \ref{fig:ptmuon}.
}
\label{fig:CMS_recast}
\end{figure}
%

%
\figref{CMS_recast} shows 
normalized distributions  of the $h\to \tau\mu\varphi$ signal as a function of the collinear mass, $m_{\tau\mu}^{\rm (Coll)}$, for all the benchmark $\varphi$ masses in each of the six signal categories. These should be compared with the generated $h\to \tau\mu$ normalized distribution, denoted by the black line. 
We see that 
the 
larger missing-energy available in $h\to\tau\mu\varphi$ decays
results in wider $m_{\tau\mu}^{\rm (Coll)}$ distributions. While the $h\to\tau\mu$ collinear mass distribution is  centered around the value of the Higgs mass, 
$m_h = 125~\rm{GeV}$, the $h\to\tau\mu\varphi$ distributions are significantly shifted, and centered at a value well below $m_h$.
Since the $h\to\tau\mu\varphi$ distributions are wide, they still contribute substantially  in the signal region, $100~\rm{GeV} \le m_{\tau\mu}^{(Coll)} \le 150~\rm{GeV}$.

\figref{ptmuon} also demonstrates that the $p_T$ distributions of the two decay topologies are well separated, and could potentially be distinguished in a future experimental analysis.

Finally, we remark that the wide $h\to\tau\mu\varphi$ signatures could potentially leak into the signal regions in $Z\to\tau\mu$ searches. The present $Z\to\tau\mu$ ATLAS search~\cite{Aad:2016blu} has reduced sensitivity to these types of decays, though, as it only targets $Z\to\tau_h\mu$,  while employing the MMC method.

\section{Flavor observables}
\label{sec:bounds}
The flavor structure of the model in Section \ref{sec:concrete} has an almost exact $U(1)_{\mu-\tau}\times U(1)_e$ symmetry. Under $U(1)_{\mu-\tau}$ the fields in the mass eigenbasis carry charges $[\tau_{L,R}]_Q=1$, $[\mu_{L,R}]_Q=1+z$, $[\varphi]_Q=z$, while $[e_{L,R}]_Q=0$. The $U(1)_e$ charge is carried only by the electron. All the $c_{ij}$ and $c_{ij}'$ couplings in \eqref{eq:Lmed} are forbidden by the $U(1)_{\mu-\tau}\times U(1)_e$ symmetry, except the $c_{23}$ and $c_{32}'$ couplings. In our model $c_{23}\sim {\mathcal O}(1)$, while from the point of view of $U(1)_{\mu-\tau}$ symmetry the $c_{32}'$ coupling is accidentally small. The $c_{23}$ can be made real through phase rotations of $\mu_L, \tau_R$. The contributions to the anomalous electric dipole moment are thus small, suppressed by small symmetry breaking terms.

The $U(1)_{\mu-\tau}\times U(1)_e$ symmetry, if exact, would forbid the flavor changing transitions that do not have $\varphi$ in the final state.
In that case $h\to \tau\mu\varphi$ is allowed, while $\ell_j\to\ell_i\gamma$ decay, $\ell_j \to\ell_i$ conversions, and $\ell_j\to\ell_i\ell_k\ell_m$ decays are forbidden. The $U(1)_{\mu-\tau}\times U(1)_e$ symmetry is broken by small nonzero entries in $c$ and $c'$ matrices, which induce at one loop level the $\ell_j\to\ell_i\gamma$ decays and $\ell_j \to\ell_i$ conversions, and from tree level exchange of $\varphi$ the  $\ell_j\to\ell_i\ell_k\ell_m$ decays.
While the neutrino sector explicitly breaks the $U(1)_{\mu-\tau}\times U(1)_e$ symmetry, its effects are suppressed by the tiny neutrino masses. For leptonic decay purposes, neutrinos effectively carry the same $U(1)_{\mu-\tau}\times U(1)_e$ charge as their corresponding leptons, and the five-body decays $\ell_j\to\ell_i\bar\ell_m\ell_m\nu_j\bar\nu_i$
are therefore allowed. The symmetry breaking effects from $c$ and $c'$ may contribute to these processes, and also allow for a more general flavor structure in these decays.

The diagram that mediates $\tau\to\mu\gamma$, $\tau\to e\gamma$ and $\mu\to e\gamma$ decays is shown in Fig. \ref{fig:1loop:dipole}. Note that the 2-loop Barr-Zee type diagrams, which in many cases give leading contributions, are always smaller for the flavor structures \eqref{eq:coupling_pattern_mass_basis}. 
The $\tau\to\mu\gamma$ decay is described by an effective Lagrangian
\beq\label{Leff}
{\cal L}_{\rm eff}=c_L Q_{L\gamma}+c_R Q_{R\gamma}, 
\eeq
where 
\beq
Q_{L\gamma,R\gamma}=\frac{e}{8\pi^2}m_\tau (\bar \mu \sigma^{\mu\nu}P_{L,R}\tau)F_{\mu\nu},
\eeq
and with obvious replacements for $\tau\to e\gamma$ and $\mu\to e\gamma$ decays. For the flavor textures in \eqref{eq:coupling_pattern_mass_basis} the dipole coefficients are dominated by
\begin{align}
c_R|_{\tau\to\mu\gamma}&=\frac{v^2}{16 \Lambda^2 m_\varphi^2} \left(\frac{1}{3}c_{23}c^*_{33}-
c_{23} c'_{22}\frac{m_\mu}{m_\tau}
\left(3 +4 \log r_\mu\right)\right),
\\
c_L|_{\tau\to e\gamma}&=\frac{v^2}{48 \Lambda^2 m_\varphi^2} 
 c_{23}c^*_{21},
\qquad
c_R|_{\tau\to e\gamma}=- \frac{v^2}{16 \Lambda^2 m_\varphi^2} 
 c_{23}c'^*_{12}
\frac{m_\mu}{m_\tau}
\left(3+4 \log r_\mu\right),
\\
c_R|_{\mu\to e \gamma}&=\frac{v^2}{16 \Lambda^2 m^2_\varphi}
\left(\frac{1}{3}c_{13}
 c_{23}^*
-
c_{13} c'_{32}\frac{m_\tau}{m_\mu}
\left(3 +4 \log r_\tau\right)
\right),~~
\end{align}
where $r_j = m_j / m_\varphi$, and we have approximated the loop functions in the limit $m_{\mu,e}\ll m_\tau \ll m_\varphi$, keeping only the leading contributions. 
The complete expression for the dipole coefficients $c_L$ and $c_R$, the full loop-functions, as well as their approximate forms, are collected in \appref{analytic_expression}.

The resulting decay widths are
\beq
\Gamma(\ell_j \to \ell_k \gamma)=\frac{\alpha m_j^5}{64 \pi^4} (|c_L|^2 +|c_R|^2),
\eeq
Numerically, this gives
\begin{align}
{\rm Br}(\tau\to \mu \gamma)&\sim 1.8 \cdot 10^{-11} \times 
\bigg|0.23\,
\frac{c^*_{33}} {\lambda^{6}}\frac{c_{23}}{ 1}+0.77 \,
\frac{c_{22}'}{\lambda^{6}}  \frac{c_{23}}{1}
\bigg|^2 \Big(\frac{\lambda}{0.2}\Big)^{12} R_{\Lambda,m_\varphi}, 
\\
{\rm Br}(\tau\to e \gamma)&\sim 1.1 \cdot 10^{-16} \times
\bigg(
0.74\, 
\frac{|c^*_{21} c_{23}|^2}{\lambda^{18}}
 \frac{0.2^2}{\lambda^2}
+
0.26\, 
\frac{|c'^*_{12} c_{23}|^2}{\lambda^{20}  }
\bigg)
\Big(\frac{\lambda}{0.2}\Big)^{20} R_{\Lambda,m_\varphi},
\\
{\rm Br}(\mu\to e \gamma)&\sim 3.1 \cdot 10^{-17} \times
\bigg|
\left(0.65\,
c^*_{23}\frac{0.2^4}{\lambda^4}
+0.35 \frac{c'_{32}}{\lambda^{4}}
\right)\frac{c_{13}}{\lambda^{10}}
\bigg|^2 \Big(\frac{\lambda}{0.2}\Big)^{28} R_{\Lambda,m_\varphi},
\end{align}
factoring out the dependence on $\Lambda$ and $m_\varphi$, 
$R_{\Lambda,m_\varphi}\equiv \big({1~{\rm TeV}}/{\Lambda}\big)^4\big({15~{\rm GeV}}/{m_\varphi}\big)^4
$.
Above we used the scaling for $c_{ij}, c_{ij}'$ from \eqref{eq:coupling_pattern_mass_basis}, and evaluated the branching fractions using the full expressions for the loop functions in \appref{analytic_expression}. 
These branching ratios are well below present experimental bounds,
see Table \ref{table:LFV_bounds} for a comparison.

A number of three-body flavor violating decays of charged leptons can be mediated by both the dipole operator and the tree level  exchange of $\varphi$. 
The three-body decays 
$\tau\to \mu\mu\mu$ 
$\tau^-\to e^-\mu^+\mu^-$ 
receive the dominant contribution from a tree level $\varphi$ exchange, while the decays, 
$\mu\to eee$, $\tau\to eee$,  
are dominated by the dipole contributions.
The decay $\tau^-\to \mu^- e^+e^-$ is a special case for which the dipole and tree-level contributions are comparable, and interference should be taken into account.
The decays $\tau^-\to e^+\mu^-\mu^-$ and $\tau^-\to \mu^+ e^- e^-$ only receive tree level $\varphi$ exchange contributions. 
The numerical values for the branching ratios are given in Table \ref{table:LFV_bounds}, with further details relegated to \appref{analytic_expression}. All of the decays are well below the present  experimental bounds.

Another potentially interesting bound is due to $\mu\to e$ conversion. At present the most stringent is the $\mu\to e$ conversion in gold, which is experimentally bounded to be $\Gamma(\mu\to e)_{\rm Au}/\Gamma_{\rm capture\,Au}<7 \cdot 10^{-13}$ at 90\% CL \cite{Bertl:2006up}. In our case the $\mu\to e$ conversion is dominated by the dipole contributions, giving
\beq
\Gamma(\mu\to e)_{\rm Au}\simeq \Big(\frac{e}{16\pi^2} c_R\big|_{\mu\to e\gamma} D\Big)^2,
\eeq
where the nuclear matrix element is $D=0.189 m_\mu^{5/2}$ \cite{Kitano:2002mt}. Taking $\Gamma_{\rm capture\,Au} = 13.07 \cdot 10^6 s^{-1}\approx 8.6\cdot 10^{-18}~\rm{GeV}$ gives
\beq
\frac{\Gamma(\mu\to e)_{\rm Au}}{\Gamma_{\rm capture\,Au} }\sim  1.2 \cdot 10^{-19} \, \bigg|
\left(0.65\,
c^*_{23}\frac{0.2^4}{\lambda^4}
+0.35 \frac{c'_{32}}{\lambda^{4}}
\right)\frac{c_{13}}{\lambda^{10}}
\bigg|^2 \Big(\frac{\lambda}{0.2}\Big)^{28} R_{\Lambda,m_\varphi}.
\eeq

We also estimated the contributions from our model to the three five-body lepton decays with measured branching ratios, $\text{Br}\left(\mu^+ \to e^+e^+e^-  \bar\nu_\mu \nu_e\right)=(3.4\pm 0.4)\times 10^{-5}$ \cite{Bertl:1985mw},  $\text{Br}\left(\tau^- \to e^-e^-e^+ \bar\nu_e \nu_\tau\right)=(2.8\pm1.5)\times10^{-5}$, and $\text{Br}\left(\tau^- \to \mu^-e^-e^+ \bar\nu_\mu \nu_\tau\right)<3.6\times 10^{-5}$ \cite{Alam:1995mt}. 
The neutrinos in the final state appear as missing-energy in the detector.  The transitions of the type 
$\ell_j \to \ell_j\ell^+_i\ell^-_i\varphi/\varphi^*$
and $\ell_j \to \ell_j\ell^+_i\ell^-_i \bar\chi\chi$
could thus contribute to the observed rates of the above five-body decays. However, since we take $m_{\varphi,\chi}>m_\tau-m_\mu$ such decays are kinematically forbidden. In our model the $\ell_j \to 3 \ell 2\nu$ decays therefore receive corrections only through off-shell $\varphi$ contributions which are always orders of magnitude below 
the SM predictions for these tree level charged current decays. 

\begin{figure}[t]
\centering
\includegraphics[width=0.5\textwidth]{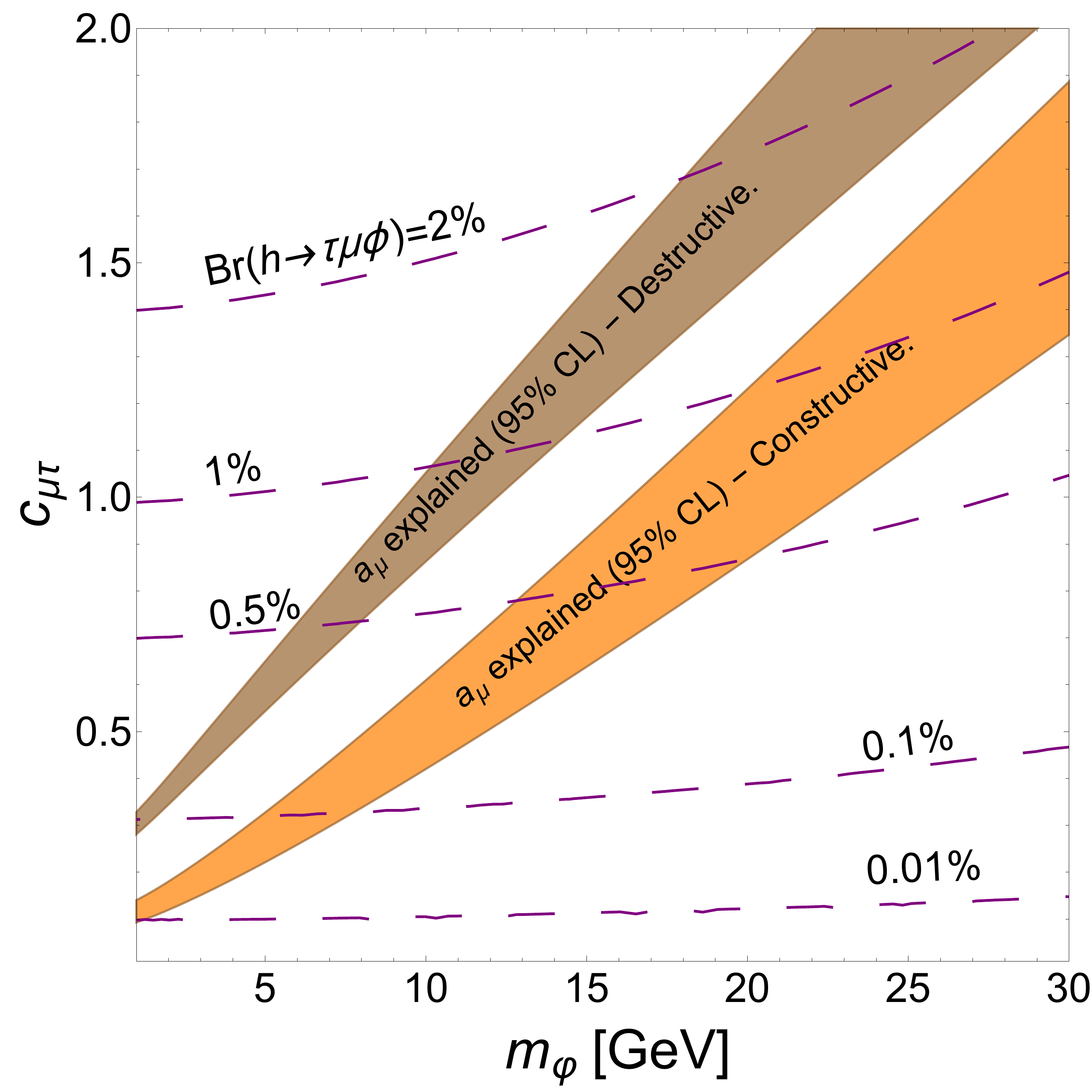}
\caption{
Contours of ${\rm Br}(h\to\tau\mu\varphi)$ in the $c_{\mu\tau}-m_\varphi$ plane, keeping other parameters fixed, are shown as dashed lines, while the orange (brown) shaded region give the  $95\%$ CL parameter sets that can explain the $(g-2)_\mu$ anomaly for the constructive (destructive) case as described in the text.
In the numerical evaluation we fixed $c'_{32} = \lambda^4$.}
\label{fig:ctaumu_vs_mphi}
\end{figure}

It is quite interesting that the flavor violating interactions of $\varphi$ can explain 
the discrepancy between measurement and SM prediction for the anomalous magnetic dipole 
moment of the muon, $a_\mu = (g-2)_\mu$,~\cite{Bennett:2006fi,Beringer:1900zz}
\beq
\Delta a_\mu = a^{\rm{experiment}}-a^{\rm{theory}} = 288(80)\times 10^{-11},
\eeq
at $95\%$ CL.
In \figref{ctaumu_vs_mphi} we show the parameter-space region in the $c_{\mu\tau}-m_\varphi$ plane which is consistent with the anomaly.
We overlay the relevant region with contours
of the $h\to\tau\mu\varphi$ branching fraction.
These are consistent with the results of the collider study (up to its embedded uncertainties), and indicate that 
$c_{\mu\tau}\sim {\cal O}(1)$
would be consistent with the observed $h\to\tau\mu\varphi$. Evidently, both phenomena could be explained in the same region of parameter-space.
 
The two dominant contributions to $\Delta a_\mu$ (see \eqref{eq:MDM}) are the scalar and pseudoscalar pieces, proportional to $|c_{23}|^2\sim 1$, and to $2\rm{Re}(c'^*_{32}c_{23})\sim 2\lambda^4$ respectively. The coupling suppression in the latter is, however, compensated by a ${m_\tau}/{m_\mu}$ enhancement and a larger integral such that the two contributions are comparable. 
The relative sign of the two contributions then determines whether they add constructively or destructively in $\Delta a_\mu$. 
This is clearly seen in \figref{ctaumu_vs_mphi} where the constructive case (orange region) requires a smaller $c_{23}$ than the destructive case (brown region).
%

In addition, the two-body decay $\tau \to \mu \bar\chi\chi$, if kinematically allowed,
would significantly change the branching ratio and the spectrum of the 3-body $\tau\to \mu \bar\nu_k\nu_j$ decays. 
This puts a constraint on the DM mass such that
$2 m_\chi>m_\tau-m_\mu$. We collect the relevant constraints on lepton flavor violating observables in~\tableref{table:LFV_bounds}.

Several other $Z$ pole measurements, while in principle sensitive to the $\varphi$ couplings, turn out to be negligible or not relevant.
The tree level  $Z\to\mu^\pm\tau^\mp\varphi/\varphi^*$ decays could potentially be captured by the LEP searches for the $Z\to\mu^\pm\tau^\mp$ decays. 
We have ${\rm Br}(Z\to\mu^\pm\tau^\mp\varphi/\varphi^*)=|c_{23}|^2\{1.5\times 10^{-5},~7.2\times10^{-6},~3.8\times10^{-6},~2\times10^{-6}\}$ for $m_\varphi = \{5,~10,~15,~20\}~\rm{GeV}$, respectively, mostly well below even the  strictest bound, ${\rm Br}(Z\to\mu^\pm\tau^\mp)<1.2\times 10^{-5}$~\cite{Abreu:1996mj}.
Furthermore, the $Z\to\mu^\pm\tau^\mp$ searches at LEP 
enforced an isolation requirement of back-to-back leptons which dramatically reduces the acceptance to the three-body decay. These searches thus do not 
constrain our model.

The $Z\bar\ell\ell$ interactions get modified at 1-loop by $\varphi$ exchanges in the vertex corrections, leading to potentially relevant universality violations in $Z\to \ell\ell$ decays, and to LFV $Z$-decays. However, these contributions are UV sensitive. The counterterm that cancels the divergence requires a dimension 6 operator in the EFT whose coefficient is otherwise not fixed.~\footnote{
The divergence is proportional to $g_L-g_R$, with $g_{L,R}$ the couplings of $Z$ to the leptons, since if SM leptons were vector-like then \Eqref{eq:lowE_varphi} could have been a renormalizable interaction.
}
 Setting it to zero gives the shifts in $g_{R\tau}$ and $g_{L\mu}$ couplings to be $\Delta g_{R\tau} \approx - 0.8 \Delta g_{L\mu} \approx |c_{23}|^2 10^{-4}$, below the sensitivity of universality measurements at LEP and SLD~\cite{ALEPH:2005ab} (see also \cite{Altmannshofer:2016brv}).

At 1-loop one also generates the $Z\to \varphi\varphi^{(*)}$ decay that contribute to $Z\to invisible$. 
However, the induced $Z\to \varphi\varphi^{(*)}$ coupling, $g_{Z\varphi\varphi^(*)} \sim \left({v}/{4\pi \Lambda}\right)^2 g_{Z\nu\bar\nu}\sim 10^{-3} g_{Z\nu\bar\nu}$, is much too small to make a noticeable effect on ${\rm Br}(Z\to invisible)=(20.00\pm0.06)\%$~\cite{Olive:2016xmw}.

\begin{table}[t!]
\centering
\begin{tabular}{lcc}
\hline\hline
LFV Process & ~Present Bound~ & ~Our Model~  \\
\hline
\underline{\textbf{Radiative Decays}} & & \\
$\rm{Br}(\mu^+ \to e^+ \gamma)$ 
& $5.7 \times 10^{-13}$ \cite{Adam:2013mnn} 
& $3.1 \times 10^{-17}$
\\
$\rm{Br}(\tau^\pm \to e^\pm \gamma)$ 
& $3.3 \times 10^{-8}$ \cite{Aubert:2009ag}
& $1.1 \times 10^{-16}$
\\
$\rm{Br}(\tau^\pm \to \mu^\pm \gamma)$ 
& $4.4 \times 10^{-8}$ \cite{Aubert:2009ag}
& $1.8 \times 10^{-11}$
\\ 
\hline
\underline{\textbf{$\mu\to e$ Conversion in Nuclei}} & & \\
$\Gamma(\mu\to e)_{\rm Au}/\Gamma_{\rm capture\,Au}$
& $7 \times 10^{-13}$ at 90\% CL \cite{Bertl:2006up}
& $1.2\times 10^{-19}$
\\ 
\hline
\underline{\textbf{3-Body Decays}} & & \\
$\rm{Br}(\mu^+ \to e^+e^+e^-)$ 
& $1.0 \times 10^{-12}$\cite{Bellgardt:1987du} 
& $D~1.9 \times 10^{-19}$
\\
$\rm{Br}(\tau^- \to \mu^- \mu^+\mu^-)$ 
& $2.1\times10^{-8}$\cite{Hayasaka:2010np} 
& 
$T~1.4\times10^{-9}$
\\
$\rm{Br}(\tau^- \to e^- e^+e^-)$ 
& $2.7\times10^{-8}$\cite{Hayasaka:2010np} 
& $D~1.1\times10^{-18}$
\\
$\rm{Br}(\tau^- \to e^- \mu^+\mu^-)$ 
& $2.7\times10^{-8}$\cite{Hayasaka:2010np} 
& 
$T~1.9\times10^{-13}$
\\
$\rm{Br}(\tau^- \to \mu^- e^+e^-)$ 
& $1.8\times10^{-8}$\cite{Hayasaka:2010np} 
& $\begin{cases}D~1.8\times10^{-13}\\T~1.9\times10^{-13}\end{cases}$
\\

\hline
$\rm{Br}(\tau^- \to e^+ \mu^-\mu^-)$ 
& $1.7\times10^{-8}$\cite{Hayasaka:2010np} 
& $T~4.9\times10^{-26}$
\\
$\rm{Br}(\tau^- \to \mu^+ e^- e^-)$ 
& $1.5\times10^{-8}$\cite{Hayasaka:2010np} 
& $T~2.1\times10^{-27}$
\\
\hline
\underline{\textbf{Muon $g-2$}} & & \\
$\Delta a_\mu$ 
& $288(80)\times 10^{-11}$ \cite{Bennett:2006fi} 
& $4.3\times 10^{-9}$
\\
\hline\hline
\end{tabular}
\caption{Charged lepton flavor violating observables
for $m_\varphi = 15~\rm{GeV}$.
The analytic expressions for the radiative decays and the dipole contributions to the 3-body decays are given in \appref{analytic_expression}.
The dipole contributions are marked by $D$, while the
tree-level ones by $T$. The tree-level contributions to the 3-body 
decays have been calculated using {\tt MadGraph~5}.
For $\Delta a_\mu$ we quote the value for $c_{23} = 1,~c'_{32}=\lambda^4$ (see discussion in the text and \figref{ctaumu_vs_mphi}).
}
\label{table:LFV_bounds}
\end{table}

\section{Dark matter phenomenology}
\label{sec:darksec}
The flavorful DM sector introduced in Section \ref{sec:model} also has interesting phenomenology. We point out several salient features of the model, while leaving the details for future work. 
The two DM states $\chi_1, \chi_2$ have the largest annihilation cross section for $\bar \chi_1 \chi_2\to \tau^+ \mu^-$ process because of the large off-diagonal couplings, $g_{12}^{L,R}$. Taking $m_{\chi_1}\simeq m_{\chi_2}\equiv m_\chi$, and $m_\chi\ll m_\varphi$ the annihilation cross section for non-relativistic $\chi_{1,2}$ is given by 
\beq
\begin{split}
(\sigma_{\rm ann} v)_{\bar \chi_1 \chi_2\to \tau^+ \mu^-} &\simeq \frac{1}{128 \pi}\frac{1}{m_\chi^2} (g_{12}^{R,L})^2\big[(c_{23})^2+(c_{23}')^2\big]\Big(\frac{v_{\rm EW}}{\Lambda}\Big)^2 \Big(\frac{m_\chi}{m_\varphi}\Big)^4.
\\
&= 4.4 \cdot 10^{-26} \frac{{\rm cm}^3}{s} \times \Big(\frac{1{\rm~TeV}}{\Lambda}\Big)^2 \Big(\frac{20{\rm ~GeV}}{m_\varphi}\Big)^4 \Big(\frac{m_\chi}{2{\rm~GeV}}\Big)^2,
\end{split}
\eeq
where in the last line we set the couplings to 1. The region of parameter space that leads to a sizable $h\to \tau\mu\varphi$ decay, can thus also have the lightest of the two $\chi_{1,2}$ a thermal relic, which requires the annihilation cross section of $\sigma_{\rm ann} v\sim 2.2 \cdot 10^{-26}{\rm cm}^3/{\rm s}$. The $\chi_{1,2}$ are kept in thermal equilibrium through flavorful annihilation, $\bar \chi_1 \chi_2\to \tau \mu$. The correct relic density is obtained with the flavor ansatz  \eqref{eq:coupling_pattern_mass_basis}, \eqref{eq:chi_varphi_cpl} for $m_\varphi \sim {\mathcal O}(10~\rm{GeV})$, and with DM mass $m_\chi \ll m_\varphi$ (the DM mass needs to be large enough, $m_{\chi}>(m_\tau+m_\mu)/2$ that the $\tau\mu$ annihilation channel is still open). If $\chi_1$ and $\chi_2$ are exactly degenerate then both states are stable and constitute DM. In general this will not be the case and the heavier state will decay. If the $\chi_2\to \chi_1 \mu^+ \mu^-$ decay channel is open, the decay will occur well before Big  Bang Nucleosynthesis with the typical decay time in milliseconds. If only $\chi_2\to \chi_1 \mu^+ e^-$, or $\chi_2\to \chi_1 e^+ e^-$, $\chi_2\to \chi_1 \gamma$ are open, however, one may run into cosmological constraints.  

DM scattering on nuclei is generated only at three loop order, from the two-loop matching onto the Rayleigh operators of the type $\bar \chi \chi F^{\mu\nu} F_{\mu\nu}$. 
Direct detection bounds are thus well below present constraints. The indirect detection signal is larger. For heavy enough DM the dominant annihilation channel would again be $\bar \chi_1 \chi_2\to \tau^+ \mu^-$. A thermal relic DM annihilating exclusively to a $\tau^+\tau^-$ final state is excluded from stacked dwarf spheroidal limits on gamma ray flux measurements by Fermi-LAT if it has mass below $m_\chi\lesssim 70$ GeV \cite{Ackermann:2015zua}. If the annihilation is to $\mu^+\mu^-$, the limits is $m_\chi\lesssim 10$ GeV \cite{Ackermann:2015tah}. For $\bar \chi_1 \chi_2\to \tau^+ \mu^-$ the limit lies between the two extremes, where the precise value would require a dedicated analysis. The limit disappears, however, if DM is asymmetric, where the $\bar \chi_1 \chi_2\to \tau^+ \mu^-$ process merely annihilates efficiently away the symmetric part of DM density.  

The above discussion changes, if one modifies the charge assignments for $\chi_{1,2}$. For instance, one could entertain different flavor charge assignment for left- and right-handed components of $\chi_{1,2}$, just as we have for the SM fields. It is then easy to arrange that the lightest state has large couplings to $\varphi$, for instance by setting $[\chi_{L1}]_Q-[\chi_{R1}]_Q=[\varphi]_Q$ and $[\chi_{L2}]_Q=[\chi_{R2}]_Q$.

\section{Conclusions}
\label{sec:conclusions}
The flavor violating Higgs decays such as $h\to \tau\mu$ can be mimicked by three-body decays of the form $h\to \tau\mu\varphi$, where $\varphi$ escapes detection and exhibits itself as an additional missing-energy. Both,  $h\to \tau\mu$ and $h\to \tau\mu\varphi$ decays, if discovered, would imply the existence of New Physics. The $h\to \tau\mu\varphi$ can dominate over $h\to \tau\mu$, if $\varphi$ carries a flavor charge, so that the $h\to \tau\mu\varphi$ decay is not flavor violating. This is in contrast to $h\to\tau\mu$ decay, which is always suppressed by a small flavor violating coupling.

In this paper we explored such a scenario where $\varphi$ is a portal to the dark sector.  In our example both $\varphi$ and the dark matter candidates carry flavor charges. The $h\to\tau\mu\varphi$ decay leads to a wider distribution in the collinear $\tau\mu$ mass, compared to the two-body $h\to\tau\mu$ decay. The two distributions are indistinguishable with present accuracy of the LHC data, if $\varphi$ is light, below about ${\mathcal O}(20{\rm~GeV})$. This can be improved with the LHC data at 13 TeV, potentially distinguishing the two scenarios. In our framework $\varphi$ couples to dark matter states with ${\mathcal O}(1)$ couplings, so that the dark matter candidate can be a thermal relic.

As a concrete example we used a $U(1)\times U(1)$ Frogatt-Nielsen flavor model and showed that there is a natural assignment of flavor charges that leads to phenomenologically acceptable range of charged lepton masses, while at the same time accounting for both the muon anomalous magnetic moment anomaly
as well as the slight $h\to\tau\mu$ excess. The relevant region of parameter-space
could be readily probed in future searches in the $13~\rm{TeV}$ dataset. The predicted rates for flavor violating processes are below present experimental bounds with the exception of the five-body lepton decays which saturate the current limits.

In the present work we refrained from modeling the neutrino mass matrix. A flavor pattern in agreement with observations can be achieved from the dimension 5 $(\bar L_i H)(\bar L_j H)$ operator. One option is a slight modification of our flavor charge assignments, that corresponds to charging the Higgs under the flavor symmetry. This would still require an additional source of electroweak symmetry breaking contributing to the  $\nu_3\nu_3$ entry of the mass matrix, in order to get the correct pattern of neutrino masses and mixings. We suspect that other options are possible and leave the details for future study.

${}$\\
\indent {\bf Acknowledgments.}
IG is supported by National Science Foundation 
grants PHY-1316792 and PHY-1620638.
JZ is supported by in part by the U.S. National Science Foundation under CAREER Grant PHY-1151392.
We would like thank
Enrique Kajomovitz, 
Arvind Rajaraman,
Yael Shadmi,
David Shih, 
Yuri Shirman,
Yotam Soreq,
Tim Tait, and
Philip Tanedo 
for useful discussions.
IG is grateful to the Mainz Institute for Theoretical Physics (MITP)
for its hospitality and its partial support during the completion of this work.

\appendix

\section{Constraints on a singlet coupled to leptons}
\label{sec:analytic_expression}

\begin{figure}[t]
\centering
\includegraphics[width=0.38\textwidth]{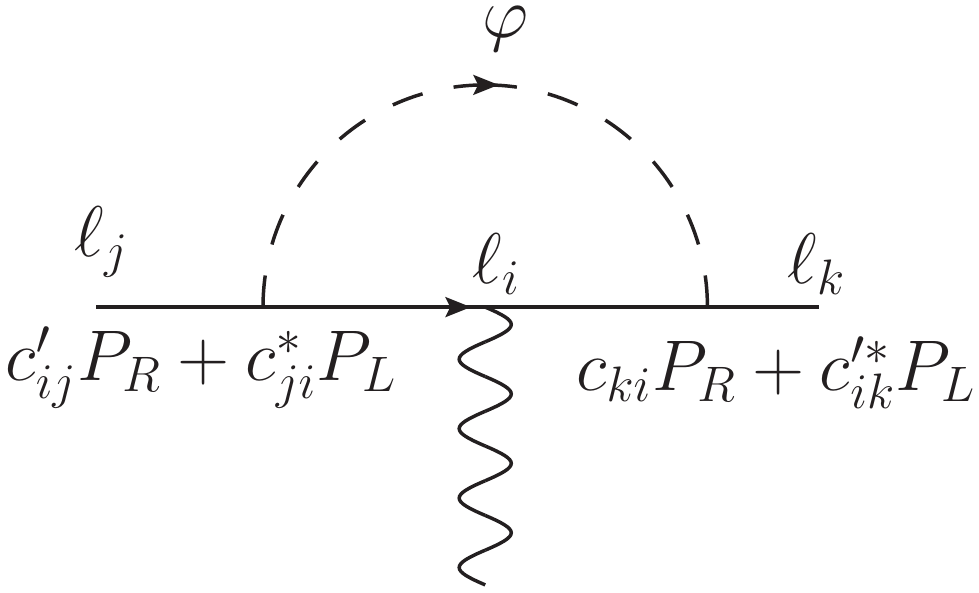}
\qquad\quad
\includegraphics[width=0.38\textwidth]{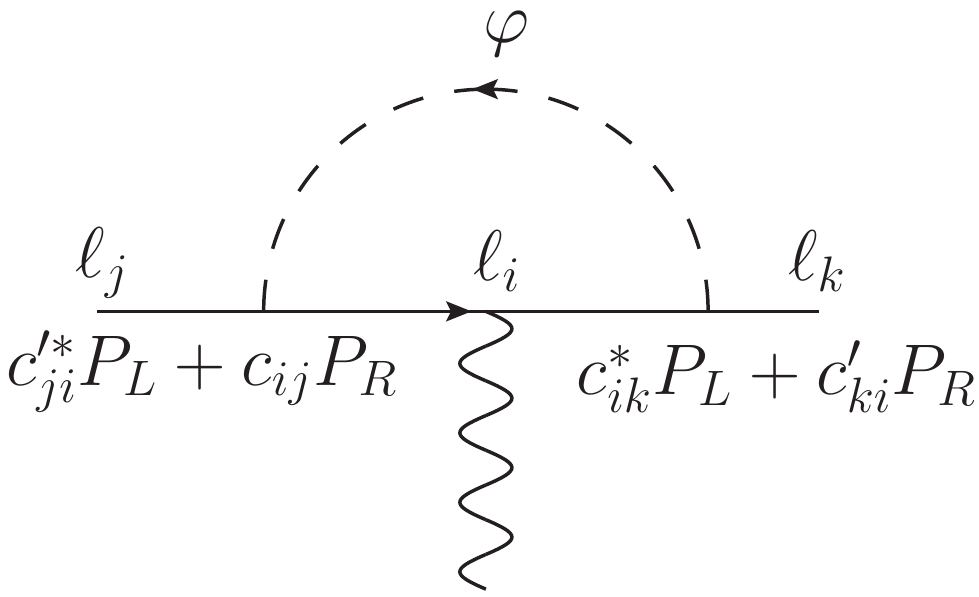}
\caption{The two $\varphi$ exchange diagrams that contribute to $\ell_i\to \ell_j\gamma$ transition at 1 loop.}
\label{fig:1loop:dipole}
\end{figure}
%

After the electroweak symmetry breaking the couplings of $\varphi$ to leptons, Eq.~\eqref{eq:Lmed}, are given by
\beq
{\cal L}_{\rm vis-med.}\supset\frac{v}{\sqrt{2}\Lambda}
\left[
\bar \ell_i \left(c_{ij} P_R  + c'^*_{ji} P_L  \right) \ell_j \varphi 
+\bar \ell_i \left(c'_{ij} P_R  + c^*_{ji} P_L  \right) \ell_j \varphi^* 
\right].
\label{eq:lowE_varphi}
\eeq
One loop $\varphi$ exchange induces the $\ell_j\to \ell_k \gamma$ transitions, see Fig. \ref{fig:1loop:dipole}, described by the dipole operators, 
\beq
{\cal L}_{\rm dipole} \supset
\frac{e}{8\pi^2}
m_j~
\bar\ell_k
\sigma^{\mu\nu}
\left(
 c^L_{kj} P_L
+
c^R_{kj} P_R 
\right)
\ell_j
F_{\mu\nu}.
\eeq
The two Wilson coefficients are given by~\cite{Harnik:2012pb}
\begin{align}
\begin{split}\label{eq:ckjL}
c^L_{kj}=&\frac{v^2}{8 m_j \Lambda^2}
\int_0^1dxdydz~\delta(1-x-y-z)\times
\\
&\qquad\times
\frac{ 
xz m_k (c'_{ki}c'^*_{ji} + c_{ki}c_{ji}^*) 
+ yz m_j( c^*_{ik}c_{ij} + c'^*_{ik}c'_{ij})
+ (x+y) m_i (c^*_{ik}c'^*_{ji} + c'^*_{ik}c_{ji}^*)
}
{
zm^2_\varphi - xz m^2_k - yz m^2_j + (x+y) m^2_i 
},
\end{split}
\\
\begin{split}\label{eq:ckjR}
c^R_{kj}=&\frac{v^2}{8m_j\Lambda^2}
\int_0^1~dxdydz~\delta(1-x-y-z)\times
\\
&\qquad\times
\frac{ 
xz m_k (c^*_{ik}c_{ij} + c'^*_{ik}c'_{ij}) 
+ yz m_j( c'_{ki}c'^*_{ji} + c_{ki}c_{ji}^*)
+ (x+y) m_i (c'_{ki}c_{ij} + c_{ki}c'_{ij})
}
{
zm^2_\varphi - xz m^2_k - yz m^2_j + (x+y) m^2_i 
}.
\end{split}
\end{align}
For later convenience we also define the rescaled Wilson coefficients
\beq
\tilde c_{L,R} 
=
4m_{j}\left(m_{j}^2 - m_{k}^2 \right) c_{L,R}.
\eeq
The 1-loop expressions \eqref{eq:ckjL}, \eqref{eq:ckjR} can then be re-expressed as
\begin{align}
\begin{split}\label{eq:tildecLkj}
\tilde c^L_{kj}= \frac{v^2}{2\Lambda^2}
\sum_i
\Big[
& m_{k}  (c'_{ki}c'^*_{ji} + c_{ki}c^*_{ji}) F_1^{j,i,k} 
+m_{j}  ( c^*_{ik}c_{ij} + c'^*_{ik}c'_{ij})
F_1^{k,i,j}
+ m_{i} (c^*_{ik}c'^*_{ji} + c'^*_{ik}c_{ji}^*) F_2^{j,i,k}
\Big],
\end{split}
\\
\begin{split}\label{eq:tildecRkj}
\tilde c^R_{kj}=\frac{v^2}{2\Lambda^2}
\sum_i
\Big[
& m_{k}  (c^*_{ik}c_{ij} + c'^*_{ik}c'_{ij}) F_1^{j,i,k}
+m_{j}  (c'_{ki}c'^*_{ji} + c_{ki}c_{ji}^*)
F_1^{k,i,j}
+ m_{i} (c'_{ki}c_{ij} + c_{ki}c'_{ij}) F_2^{j,i,k}
\Big],
\end{split}
\end{align}
where the loop functions are
\begin{align}
F_1^{j,i,k} = &
\frac12-
\int^1_0 dx
\frac{\Delta}{\left(m_{j}^2-m_{k}^2\right)x}
\ln
\left[
1+x(1-x){\left(m_{j}^2-m_{k}^2\right)}{\Delta}^{-1}
\right],
\\
F_2^{j,i,k} =&
\int^1_0 dx
\frac{1-x}{x}
\ln
\left[
1+x(1-x){\left(m_{j}^2-m_{k}^2\right)}{\Delta}^{-1}
\right],
\end{align}
with
\beq
\Delta = x m^2_\varphi +m_{i}^2 (1-x)  - m_{j}^2 x(1-x).
\eeq
Taking the limit where $m_k \ll m_j$ these functions can be approximated by,
\beqa
F_1^{j,i,k} &=& \frac{r_j^2 \left(r_i^6-6 r_i^4+3 r_i^2+6 r_i^2 \log \left(r_i^2\right)+2\right)}{12 \left(r_i^2-1\right)^4},
\\
F_2^{j,i,k} &=& \frac{r_j^2 \left(r_i^4-4 r_i^2+2 \log \left(r_i^2\right)+3\right)}{2 \left(r_i^2-1\right)^3},
\eeqa
where $r_j = m_j / m_\varphi$. In terms of \eqref{eq:tildecLkj}, \eqref{eq:tildecRkj} the $\ell_j\to \ell_k \gamma$ partial decay width is given by
\beq
\Gamma(\ell_j \to \ell_k \gamma)
=
\frac{\alpha}{4(4\pi)^4} 
\frac{
m_{j}^2 - m_{k}^2
}
{ m_{j}^3 }
\left(
|\tilde c_{kj}^L|^2
+|\tilde c_{kj}^R|^2
\right)\simeq \frac{4 \alpha}{(4\pi)^4} 
m_j^5
\left(
| c_{kj}^L|^2
+| c_{kj}^R|^2
\right),
\eeq
where in the last equality we assumed $m_k\ll m_j$.

If the $\ell_j \to \ell_k \bar\ell_k \ell_k$ is dominated by the dipole contribution the partial decay widths are given by \cite{Harnik:2012pb}
\begin{align}
\Gamma(\ell_j \to 3 \ell_k )&= \frac{\alpha^2 m_{j}^5}{6(2\pi)^5}
\left(
\log\frac{m_{j}^2}{m_{k}^2}-\frac{11}{4}
\right)
\frac{\left(
\left|
\tilde c_L\right|^2
+
\left|
\tilde c_R \right|^2
\right)}{16 m_j^6},
\\
\Gamma(\ell_j^- \to \ell_i^- \ell_k^+\ell_k^- )&= \frac{\alpha^2 m_{j}^5}{6(2\pi)^5}
\left(
\log\frac{m_{j}^2}{m_{k}^2}-3
\right)
\frac{\left(
\left|
\tilde c_L\right|^2
+
\left|
\tilde c_R \right|^2
\right)}{16 m_j^6},
\end{align}
where we used that $m_j\ll m_k$ and kept only the leading terms from the phase space integral.

The anomalous magnetic moment we express as
\beq
a_{\ell_j} = 
\frac{m_{j}}{16\pi^2}
\sum_i\int_0^1
dx (1-x)^2
\frac{ 
x m_{j} S^{(j)}_i
+
m_{i} P^{(j)}_i
}
{
x m^2_{\varphi} + (1-x) m^2_{i} - x(1-x) m^2_{j}
},
\label{eq:MDM}
\eeq
where
\beqa
 S^{(j)}_i &=& \frac{v^2}{2\Lambda^2}
\left(
c^*_{ij}c_{ij} + c'^*_{ij}c'_{ij} +
c^*_{ji}c_{ji} + c'^*_{ji}c'_{ji}
\right),
\\
P^{(j)}_i &=&\frac{v^2}{2\Lambda^2}
\left(
c'^*_{ji}c_{ij} + c^*_{ij}c'_{ji} +
c'^*_{ij}c_{ji} + c^*_{ji}c'_{ij}
\right).
\eeqa

Finally, we give expressions for the Higgs partial decay widths.
The Lagrangian terms in \eqref{eq:Lmed} and \eqref{eq:lepton_yukawa} give after electroweak symmetry breaking 
\beq
{\cal L} \supset - \frac{Y_{ij}}{\sqrt2}h \bar\ell_iP_R \ell_j 
+  \frac{c_{ij}}{\sqrt2\Lambda} h \bar\ell_i P_R \ell_j \varphi
+ \frac{c'_{ij}}{\sqrt2\Lambda}  h \bar\ell_i P_R \ell_j \varphi^*
+ {\rm h.c.}.
\eeq
The partial width for the two-body decay $h\to\tau\mu$ in the limit 
$m_h \gg m_\tau,m_\mu$ is given by
\beq
\Gamma(h\to\tau^\pm\mu^\mp)
=
\left( |Y_{\tau\mu}|^2+|Y_{\mu\tau}|^2\right)
\frac{m_h}{16\pi}.
\label{eq:higgs_2body}
\eeq
The partial width for the three-body decay $h\to\tau\mu\varphi$ is given by
\beqa
\label{eq:higgs_3body}
\Gamma(h\to\tau^\pm\mu^\mp\varphi~ / ~\varphi^*)
&=&
\left(
\left|c_{\tau\mu}\right|^2
+\left|c_{\mu\tau}\right|^2
+\left|c'_{\tau\mu}\right|^2
+\left|c'_{\mu\tau}\right|^2
\right)
\frac{m_h}{3(8\pi)^3}\frac{m_h^2}{\Lambda^2}
f(m_\varphi^2 / m_h^2),
\eeqa
where the phase space function is 
\beq
f(r)=(1-r)(1+r (10+r))+6 r (1+r)\ln r.
\eeq
Note that $f(0)=1$, so that the phase space factor for $m_\varphi\ll m_h$ approaches unity.

\bibliography{htaumu_phi}

\end{document}